\documentclass[prd, aps, nofootinbib, preprintnumbers, showpacs, superscriptaddress,twocolumn]{revtex4}

\usepackage{graphics,graphicx,rotating}
\usepackage[usenames]{color}
\usepackage[percent]{overpic}
\usepackage{mathrsfs,amsmath,amssymb}
\usepackage[usenames]{color}

\newcommand{\be}{\begin{equation}}
\newcommand{\ee}{\end{equation}}
\newcommand{\beq}{\begin{equation}}
\newcommand{\eeq}{\end{equation}}
\newcommand{\ber}{\begin{eqnarray}}
\newcommand{\eer}{\end{eqnarray}}
\newcommand{\bea}{\begin{eqnarray}}
\newcommand{\eea}{\end{eqnarray}}

\newcommand{\dt}{{\rm d}t}

\newcommand{\dtheta}{{\rm d}\theta}
\newcommand{\dphi}{{\rm d}\varphi}

\newcommand{\hc}{{\sf h}}

\newcommand{\AEI}{\affiliation{Max-Planck-Institut f\"ur
    Gravitationsphysik, Am M\"uhlenberg 1, 14475 Potsdam, Germany}}

\newcommand{\caltech}{\affiliation{Theoretical Astrophysics 130-33,
    California Institute of Technology, Pasadena, California 91125,
    USA}}

\newcommand{\PSU}{ \affiliation{Center for Gravitational Wave
    Physics, Pennsylvania State University, University Park,
    Pennsylvania 16802, USA}} 

\newcommand{\GT}{\affiliation{Center for Relativistic
    Astrophysics, School of Physics, Georgia Institute of
    Technology, Atlanta, Georgia 30332-0430, USA}}

\newcommand{\goddard}{\affiliation{Gravitational Astrophysics Laboratory, 
NASA Goddard Space Flight Center, 8800 Greenbelt Rd., Greenbelt,
Maryland 20771, USA}} 

\newcommand{\baltimore}{
\affiliation{Center for Space Science \& Technology, Physics
  Department, University of Maryland Baltimore County, 1000 Hilltop
  Circle, Baltimore, Maryland 21250, USA}} 

\newcommand{\cscamm}{
\affiliation{Center for Scientific Computation and Mathematical
  Modeling, University of Maryland, 4121 CSIC Bldg. 406, College Park,
  Maryland 20742, USA}} 

\newcommand{\cornell}{\affiliation{Center for Radiophysics and Space
    Research, Cornell University, Ithaca, New York 14853, USA}} 

\newcommand{\jena}{\affiliation{Theoretical Physics Institute,
    University of Jena, 07743 Jena, Germany}} 

\newcommand{\palma}{\affiliation{Departament de F\'isica, 
  Universitat de les Illes Balears, Cra.\ Valldemossa Km.\ 7.5, Palma 
de Mallorca, E-07122 Spain}}

\newcommand{\cork}{\affiliation{Physics Department, University College
    Cork, Cork, Ireland}}

\begin{document}

\title{The Samurai Project: verifying the consistency of
  black-hole-binary waveforms for gravitational-wave detection}

\author{Mark Hannam}        \cork
\author{Sascha Husa}        \palma
\author{John G. Baker}      \goddard
\author{Michael Boyle}      \caltech\cornell
\author{Bernd Br\"ugmann}   \jena
\author{Tony~Chu}           \caltech
\author{Nils Dorband}       \AEI
\author{Frank Herrmann}     \cscamm\PSU
\author{Ian Hinder}         \AEI\PSU
\author{Bernard J. Kelly}   \goddard
\author{Lawrence~E.~Kidder} \cornell
\author{Pablo Laguna}       \GT\PSU
\author{Keith D. Matthews}  \caltech
\author{James R. van Meter} \goddard\baltimore
\author{Harald P. Pfeiffer} \caltech
\author{Denis Pollney}      \AEI
\author{Christian Reisswig} \AEI
\author{Mark A. Scheel}     \caltech
\author{Deirdre Shoemaker}  \GT\PSU

\date{\today}

\begin{abstract}
We quantify the consistency of numerical-relativity black-hole-binary
waveforms for use in gravitational-wave (GW) searches with current and planned
ground-based detectors. We compare previously
published results for the $(\ell=2,\vert m \vert =2)$ mode of the
gravitational waves from an equal-mass nonspinning binary, calculated
by five numerical codes. We focus on
the $1000M$ (about six orbits, or 12 GW cycles) before the peak of the
GW amplitude and the subsequent ringdown. We find that
the phase and amplitude agree within each code's uncertainty
estimates. The mismatch between the $(\ell=2,\vert m\vert =2)$ modes is 
better than $10^{-3}$ for binary masses above $60\,M_{\odot}$ with
respect to the Enhanced LIGO detector noise curve, and for masses
above $180\,M_{\odot}$ with respect to Advanced LIGO, Virgo and Advanced 
Virgo. Between the waveforms with the best agreement, the mismatch is
below $2 \times 10^{-4}$. We find that the waveforms would be
indistinguishable in all ground-based detectors (and for the masses we
consider) if detected with a signal-to-noise ratio of less than
$\approx14$, or less than $\approx25$ in the best cases.  
\end{abstract}


\maketitle

\section{Introduction}

Direct detection of gravitational waves is expected in the next few
years by a network of ground-based laser-interferometric detectors,
LIGO \cite{Abbott:2007kv,Waldman06,LIGO_web}, Virgo
\cite{Acernese2006,VIRGO_web} and GEO \cite{GEO1,GEOStatus:2006,GEO_web},
which operate in the frequency range $\sim10^1$-$10^4$\,Hz. 
The scientific scope of gravitational-wave
observations will be extended (see \cite{Hughes:2007xm} for a
recent overview) by space-based instruments such as LISA
\cite{LISA1,Danzmann:2003tv}, which will be sensitive to signals at
significantly lower frequencies. A likely source for the first
detection, and an essential part of the science objectives of all
gravitational-wave detectors, is the merger of black-hole-binary
systems. Detection of gravitational-wave events and their further
analysis rely on the theoretical modeling of waveforms. 
Until recently, theoretical waveforms for the coalescence of black
holes were based on analytic approximations to the full general theory
of relativity, in particular the post-Newtonian expansion, which
models the signal from the slow inspiral of the two black holes, and
black-hole perturbation theory, where the complex ringdown frequencies
of black holes can be computed (see \cite{Blanchet02,Kokkotas99a}
for reviews). These methods cannot currently model from
first principles the merger phase, when the wave amplitude peaks. 
Correspondingly, data
analysis methods so far had to be developed without information from
complete black-hole-binary waveforms.  

The situation changed with breakthroughs in numerical relativity in 2005
\cite{Pretorius:2005gq,Campanelli:2005dd,Baker:2005vv}  
that made it possible to calculate the late inspiral, merger and
ringdown of a black-hole-binary system in  
full general relativity, and to calculate the gravitational waves
produced in the process. Since that time many more numerical simulations
have been performed
\cite{Diener:2005mg,Campanelli:2006fy, Campanelli:2006fg,
  Campanelli:2006uy, Campanelli:2006gf, Scheel:2006gg,
  Gonzalez:2006md, Baker:2006kr, Baker:2006ha, Baker:2006vn,
  Baker:2006yw, Sperhake:2006cy,Marronetti:2007wz,Lindblom:2007xw,
  Boyle:2007ft,Pfeiffer:2007yz, Tichy:2007hk, Hinder:2007qu, Herrmann:2007zz, Herrmann:2007ex, Vaishnav:2007nm, Schnittman:2007ij, Buonanno:2007pf, Pan:2007nw, Baker:2007gi, Choi:2007eu,Lousto:2007db,Campanelli:2007cga,
Campanelli:2007ew,Berti:2007nw,Sperhake:2007gu,Rezzolla:2007rz,Damour:2007vq,Rezzolla:2007xa,Pollney:2007ss,Koppitz:2007ev,Hannam:2007wf,Brugmann:2007zj,Hannam:2007ik,Husa:2007rh,Husa:2007hp,Gonzalez:2007hi,Damour:2008te,Brugmann:2008zz,Scheel:2008rj,Boyle:2008ge,Tichy:2008du,Healy:2008js,Hinder:2008kv,Shoemaker:2008pe,Washik:2008jr,Baker:2008mj,Baker:2008md,Campanelli:2008nk,Lousto:2008dn,Dain:2008ck,Sperhake:2008ga,Cao:2008wn,Gonzalez:2008bi,Walther:2009ng,Shibata:2008rq},
and efforts have been made to produce waveform templates based on numerical results
\cite{Buonanno:2007pf,Ajith:2007qp,Ajith:2007kx,Ajith:2007xh}.
Some of these template banks are already available to be used for
searches within the LSC Algorithm Library \cite{LAL}. There is also
an ongoing project to test search pipelines with injections of numerical  
data into simulated LIGO and Virgo noise, the Numerical INJection Analysis
(NINJA) project  \cite{Aylott08}. The work described in this paper
was conceived as complementary to that in the NINJA effort, and has
subsequently been dubbed the Samurai project.

If waveform template banks based (at least in part) on numerical results 
are to be confidently used in detector
searches, it is important to know the accuracy of the input numerical
waveforms. Most numerical waveforms are published with some internal error
analysis and uncertainty estimates. However, our goal in this paper is
to perform a stronger consistency check by comparing the results of
{\it different} numerical codes, produced using different formulations
of the Einstein equations, initial data, gauge conditions, and
numerical techniques. First studies comparing numerical waveforms 
were performed in \cite{Baker:2007fb,Bruegmann:2006at}; our project
extends that earlier work.

This paper serves two purposes: (1) to verify that the numerical
waveforms we compare agree with each other within the uncertainty
estimates originally published with those waveforms, and (2) to
quantify the differences between the waveforms in terms and  
measures meaningful to both the numerical-relativity and
gravitational-wave data-analysis communities. In particular, in
addition to making a direct comparison between the phase and amplitude
of the respective waveforms, we also compute their mismatch 
with respect to the Enhanced LIGO, Advanced LIGO,  Virgo and 
Advanced Virgo detectors,
and the maximum signal-to-noise ratio (SNR) below which these waveforms 
would be indistinguishable. 

In the present comparison we focus on the physical system that has
been studied in most detail by the numerical-relativity community: a
binary consisting of equal mass, non-spinning black holes following
non-eccentric inspiral and merger.  Specifically, we consider the dominant
$(\ell=2, \vert m \vert =2)$ spherical-harmonic mode,
which has been the focus of most data-analysis research to date, and
is the most important from the point of view of GW detection.
It is now known that
the sub-dominant modes are also important for parameter estimation,
but we will leave an anlysis of those to future work; see also our
comments in the Conclusion.
In order to keep the data analysis
aspects of this paper straightforward, we will discuss
only single detectors, and neglect
the subtleties introduced when dealing with networks of detectors or with
time-delay interferometry.

We will consider the waveform
from roughly $1000M$ before merger, to about $80M$ after merger, where
$M$ is the total binary mass in geometrical units. This is about six
orbits before merger, or $0.005\,(M/M_{\odot})$ seconds, where
$M_{\odot} = 1.477 \times 10^{3}$\,m is the mass of the sun.

We make use of results from the 
{\tt BAM} \cite{Bruegmann:2006at,Husa2007a}, {\tt CCATIE}
\cite{Pollney:2007ss}, {\tt Hahndol} \cite{Imbiriba:2004tp,vanMeter:2006vi} and 
{\tt MayaKranc} codes \cite{Vaishnav:2007nm}, which all use the 
BSSN/moving-puncture
\cite{Campanelli:2005dd,Baker:2005vv,Shibata95,Baumgarte1999,Alcubierre02a,vanMeter:2006vi,Hannam:2006vv,Gundlach2006}   
approach and finite-difference techniques, and the {\tt SpEC} 
code \cite{Scheel:2006gg},
which solves a variant of the generalized-harmonic system
\cite{Friedrich:2000qv,Lindblom:2005qh,Lindblom:2007xw}
using pseudospectral methods.  

The paper is organized as follows. In Section~\ref{sec:waveforms} we
summarize the numerical waveforms that we analyze, and the codes that
were used to produce them. In Section~\ref{sec:phase} we directly
compare the phase and amplitude of the waveforms. In
Section~\ref{sec:match} we calculate the detector mismatch between the
waveforms for a range of masses and detectors, and determine the SNR 
below which the waveforms would be indistinguishable for a single GW
detector. In Section~\ref{sec:discuss} we draw some conclusions from our
comparisons.

\section{Numerical waveforms}
\label{sec:waveforms}

\subsection{The physical system} 

We restrict our attention to the modeling of one physical system, the
orbital inspiral of two 
black holes of {\em equal mass and zero spin} with vanishing eccentricity. 
The size of the orbits decreases as the system loses energy through
gravitational radiation emission, until the black holes merge to form
a single spinning black hole. The final mass and spin, which determine
for example the ringdown frequencies, are routinely determined from
numerical simulations with a variety of methods, for early results
see
e.g.~\cite{Baker2002,Alcubierre2003:pre-ISCO-coalescence-times,Pretorius:2005gq,Campanelli:2005dd,Baker:2005vv}.  
Current simulations are able to obtain very
accurate results for the final black hole parameters, and as examples
we quote the results and error estimates 
reported for the {\tt SpEC} and {\tt BAM} codes 
\cite{Scheel:2008rj,Damour:2008te}, which are consistent within
the given error estimates:
the mass of the final black hole has been found as 
$M_f = 0.95162 \pm 0.00002 M$ with {\tt SpEC}, and as 
$M_f = 0.9514  \pm 0.0016  M$ with {\tt BAM};
the dimensionless spin (Kerr parameter) has been computed as
$S_f /M_f^2 = 0.68646 \pm 0.00004$ with {\tt SpEC} and as 
$S_f /M_f^2 = 0.687   \pm 0.002$   with {\tt BAM}. This corresponds to
a dominant ringdown frequency of the $\ell=m=2$ mode of
$M \omega = 0.5539 \pm 0.0018 $  (conservative {\tt BAM} estimate) or
$M \omega = 0.5535 \pm 0.00003$ ({\tt SpEC} estimate), using interpolation
in tabulated values of ringdown frequencies given in \cite{Berti:2005ys}.

If we decompose the gravitational-wave
signal from this system into spherical harmonics, the $(\ell=2,m=\pm 2)$
modes dominate. The frequency of these two modes (which are related by
a $\pi/2$ phase shift) is very close to twice that
of the orbital motion during inspiral, and steadily increases as the
black holes approach merger. The amplitude of the signal is a function
of the frequency, and also increases. The signal frequency and
amplitude peak around merger, and then the amplitude decays
exponentially as the merged black hole rings down to a stationary Kerr
black hole. For an equal-mass nonspinning binary, we know from
numerical simulations that the peak frequency is approximately 
$f_{peak} \approx 16 (M_{\odot}/M)$\,kHz. Five orbits before merger, 
the wave frequency is $f \approx 1.95 (M_{\odot}/M)$\,kHz, and one 
hundred orbits before merger the frequency is 
$f \approx 0.38 (M_{\odot}/M)$\,kHz, as estimated by post-Newtonian methods. 

The most sensitive ground-based detectors currently in operation, LIGO
and Virgo, can detect signals from black-hole binary (BH-BH)
mergers out to distances
of up to several hundred Mpc (depending on the binary's mass and
orientation, see e.g.,~\cite{Abbott:2007xi,Abbott:2007kv,Ajith:2007kx}). 
Estimated event rates for BH-BH coalescence events are
on the order of one every few years, but with large uncertainties,
see
e.g.~\cite{Cutler:2002me,Kalogera:2006uj,Belczynski:2006zi,Sadowski:2007dz}.  
The future Enhanced LIGO \cite{Adhikari06} and Virgo+ \cite{Acernese2006}
detectors will increase event rates by $\sim5$ times, 
whereas the Advanced LIGO \cite{AdvLIGO} and Advanced Virgo \cite{Flaminio05}
detectors will increase event rates by roughly three orders of
magnitude as compared with the current detectors. 

\begin{figure}[t]
\centering
\includegraphics[width=85mm,bb=0 0 480 295]{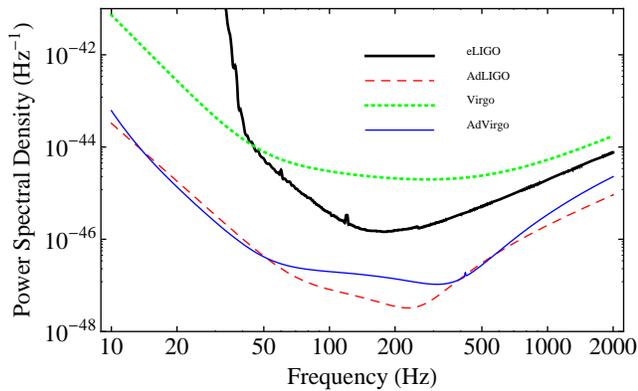}
\caption{Theoretical noise curves (power spectral density $S_n$, see 
Eq.~\ref{eq:scalar_prod}) for the detectors Enhanced LIGO,
  Advanced LIGO, Virgo and Advanced Virgo.}
\label{fig:noise}
\end{figure}

These detectors are sensitive to frequencies
ranging from $\sim 10-40$~Hz up to $\sim 2$~kHz. 
The merger signal will be in this frequency range for systems with
total masses of roughly 5--250\,$M_{\odot}$.  
The merger will be in the most sensitive part of the detectors'
frequency bands for masses around 50\,$M_\odot$, and for that case 
the detectors will also be sensitive to the signal from the last ten
orbits before merger. Theoretical estimates of the noise curves for
the four detectors we consider, Enhanced LIGO,
Advanced LIGO \cite{LAL},
Virgo \cite{LAL} and Advanced VIRGO \cite{Flaminio05}, 
are shown in Figure~\ref{fig:noise} (for Virgo and Advanced LIGO we
use approximate analytical 
formulas as displayed in \cite{Ajith:2007kx}).

Astrophysical black holes may form binaries through a number of mechanisms
\cite{Kalogera:2006uj,Postnov:2007jv,lrr-2006-2}. In general the black holes
will have different masses and will be spinning (high spins may be 
typical~\cite{Volonteri:2004cf,Gammie:2003qi,Shapiro05}),
and the orbits will be eccentric. But gravitational-radiation emission reduces 
the eccentricity, so for typical comparable-mass inspirals the eccentricity is
expected to be negligible \cite{Peters:1964} by the
time the binaries have reached the frequencies we have just
discussed (for the situation
in globular clusters see however \cite{lrr-2006-2}). 
For this reason most analytical and
numerical work in modelling gravitational-wave signals has focussed on
binaries that follow non-eccentric (or ``quasi-circular'') inspiral;
see however
\cite{Sperhake:2007gu,Hinder:2007qu,Hinder:2008kv,Healy:2008js} for
numerical results on eccentric binaries.  

\begin{figure}[t]
\centering
\includegraphics[width=80mm]{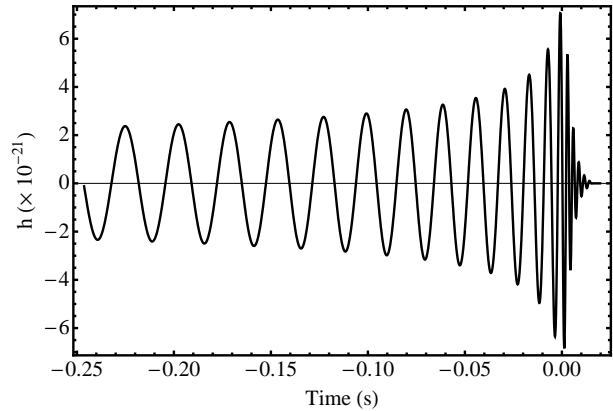}
\caption{The gravitational-wave strain from an 
optimally-oriented 60\,$M_{\odot}$ 
equal-mass nonspinning black-hole binary located 100\,Mpc away from the 
detector. The waveform covers about six orbits, or twelve GW cycles,
before merger.}
\label{fig:strain}
\end{figure}

The preceding discussion motivates a focus on the last orbits before
merger of binaries following non-eccentric inspiral. We study a binary
that consists of black holes with equal mass and no spin, simply
because this configuration has been studied in the most detail in
numerical-relativity simulations. We consider the gravitational-wave
signal from the last $\sim6$ orbits and merger of this system. 
Figure~\ref{fig:strain} shows one polarization of the gravitational-wave 
strain from an example of such a binary, with total mass 60\,$M_{\odot}$,
 optimally oriented to the detector and located 100\,Mpc away.

\subsection{Numerical codes}

To calculate the gravitational-wave signal in full general relativity,
we first require a solution of Einstein's equations. This must be
produced numerically, and can be done in a number of different ways. We
will compare the results of five computer codes: {\tt BAM}, {\tt CCATIE},
{\tt Hahndol}, {\tt MayaKranc} and {\tt SpEC}. These differ in
their procedure for constructing black-hole-binary initial data,
decomposition of Einstein's equations into a numerically well-posed
and stable form, numerical techniques used to evolve the data,
choice of gauge conditions during the evolution, and details of
the calculation of the gravitational-wave signal.

We will summarize the different methods of setting up the initial
data, the formulations of Einstein's equations, and the numerical
techniques. The purpose is not to provide a full exposition of these
methods (the full technical details can be found in the references 
given in Table~\ref{tab:codes})
but to make clear the similarities and differences of the five
codes. 

\subsubsection{Initial data}

Four of the codes, {\tt BAM}, {\tt CCATIE}, {\tt MayaKranc} and {\tt
  Hahndol}, use Bowen-York  
puncture initial data \cite{Brandt97b}. The chief features of these
data are that the spatial metric is conformally flat, and the physical
momenta and spins of the black holes can be specified directly as
parameters in the Bowen-York solution of the momentum constraint
\cite{Bowen1980}. The black holes manifest themselves in the data
through topological wormholes, which also allow the spatial slices to
bypass the black-hole singularities. The wormholes are compactified so
that their ends are mapped to single points, or ``punctures''
\cite{Beig94,Beig:1994rp,Dain01a,Brandt97b,Ansorg:2004ds}. 
The nature of these wormholes
changes during a dynamical evolution
\cite{Hannam:2006vv,Hannam:2006xw,Hannam:2008sg,Brown:2007tb}, but the
spatial slices never reach the singularities, and the data can be
evolved without recourse to excising any region of the computational domain. 

The data used in each code differ only in the choices of initial
separation of the black-hole punctures, and their momenta, and in the
method of numerically solving the Hamiltonian constraint in the
puncture approach; this last distinction will affect the accuracy of 
the solution of the Hamiltonian constraint, but we assume in this
work that the solution used by all four codes is sufficiently accurate
that the remaining numerical errors do not contribute to the differences
we measure in the final results.

Ideally the momenta are chosen to produce non-eccentric quasi-circular
inspiral. In the simulation from the {\tt Hahndol} code, the initial momenta 
were modified by hand until a roughly quasi-circular inspiral was
obtained. The {\tt BAM}, {\tt CCATIE}, and {\tt MayaKranc} simulations
used parameters 
calculated by post-Newtonian methods, as outlined in
\cite{Husa:2007rh}. This procedure results in an eccentricity of $e <
0.0016$. The choices of initial momenta are given in Table~\ref{tab:codes}.

The {\tt SpEC} code uses excision data: the data extend only to the
black-hole apparent horizons. The data were constructed by solving the
conformal-thin-sandwich initial-value equations~\cite{York99,Pfeiffer:2002iy}, 
with suitable
boundary conditions on the apparent horizons and at the outer boundary
to produce non-spinning black holes 
\cite{Cook:2001wi,Cook:2004kt,Caudill:2006hw} in an orbit with small radial
velocity~\cite{Pfeiffer:2007yz}. The parameters 
appropriate to quasi-circular inspiral were first predicted by the
methods described in \cite{Caudill:2006hw}, and then modified using
the iterative procedure described in~\cite{Pfeiffer:2007yz,Boyle:2007ft}, 
to yield
an eccentricity of below $e \sim 5\times 10^{-5}$.  Once again the data are
conformally flat.  

The reader should not make too much (or too little) of the
differences between these two types of data, Bowen-York-puncture and
conformal-thin-sandwich-excision. Although they are constructed in quite
different ways, both sets of data are based on similar choices of the
free data in initial-value equations (in particular conformal
flatness), and may not be physically very different; elucidating their
exact differences is not trivial. Conversely, they are not identical,
and there is no reason to expect {\it a priori} that the waveforms
resulting from evolutions of both sets of data will precisely
agree. Evaluating that difference is part of the analysis in this work.

\subsubsection{Evolution systems}

The codes that start with puncture initial data --- {\tt BAM}, {\tt
  CCATIE}, {\tt MayaKranc}  
and {\tt Hahndol} --- evolve the data with the BSSN formulation of
Einstein's equations, and follow \cite{Campanelli:2005dd,Baker:2005vv} in
the use of coordinate conditions that allow the black holes to move
across the grid. The BSSN evolution system
\cite{Shibata95,Baumgarte1999,Alcubierre02a} is combined with
hyperbolic evolution equations for the lapse and shift (1+log slicing
\cite{Bona97a} and the $\tilde\Gamma$-driver shift condition
\cite{Alcubierre02a,vanMeter:2006vi}), which have been shown to lead to a
well-posed initial-value problem \cite{Gundlach2006}. 

The {\tt SpEC} code uses a first-order formulation
\cite{Lindblom:2005qh} of the generalized-harmonic-gauge system
\cite{Friedrich85,Friedrich:2000qv} with built-in constraint-damping
terms \cite{Gundlach2005:constraint-damping,Pretorius2005}.  This
system is manifestly symmetric hyperbolic and well-posed. The gauge
source functions are chosen to be constant in a comoving frame during
the inspiral \cite{Scheel:2006gg}, and are evolved
according to a sourced wave equation during merger and ringdown
\cite{Scheel:2008rj}.  The characteristic fields of the system are all
outward-flowing (into the holes)
at the excision boundaries, so no boundary conditions
are needed or imposed there.  The outer boundary conditions
\cite{Lindblom:2005qh,Rinne:2006vv,Rinne:2007ui} are designed to
prevent the influx of constraint
violations~\cite{Stewart98,Friedrich99,Bardeen02,
  Szilagyi02b,Calabrese02c,Szilagyi02a,Kidder2005:boundary-conditions}
and undesired incoming gravitational radiation~\cite{Buchman:2006xf},
while allowing the outgoing gravitational radiation to pass through
the boundary.

\subsubsection{Numerical techniques}

The moving-puncture codes solve the partial-differential equations 
of the BSSN formulation of the Einstein equations with
finite-difference methods. The numerical 
domain consists of nested Cartesian domains, such that successive levels
of refinement are placed both around the individual black holes 
(and centered on the punctures) and around the entire
black-hole-binary system (centered on the origin of coordinates). The
details of 
the mesh refinement differ between codes, and the full details can be
found in the relevant references. The spatial finite-differencing
is in general fourth-order in {\tt Hahndol}, {\tt CCATIE}, and {\tt MayaKranc},
and  sixth-order in {\tt BAM}. Here, centered differences are used with
the exception of shift advection terms, which use one-point lopsided
stencils. Integration forward in time is performed with a fourth-order
Runge-Kutta method. The {\tt Hahndol} code uses a uniform time step, 
while the other BSSN codes use variants of a Berger-Oliger scheme,
where finer grids can evolve with smaller time steps. The refinement
boxes that are centered around the black holes move with them through the grid. 

The {\tt SpEC} code uses multidomain pseudospectral methods on a grid
with two excised regions, one just inside the apparent horizon of each
hole, and employs a dual-frame technique to track the motion of the
holes \cite{Scheel:2006gg}.  The computational domain consists of
two sets of concentric spherical shells, one surrounding each excised region,
another set of concentric spherical shells extending to the outer
boundary, and a structure of touching cylinders that fills in the
remaining volume and overlaps some of the spherical shells.
Inter-domain boundary conditions are enforced with a penalty
method~\cite{Gottlieb2001,Hesthaven2000}. Time stepping is
accomplished via the method of lines, using an adaptive fourth/fifth
order Runge-Kutta method.

\subsubsection{Summary of the numerical codes and waveforms}

Table~\ref{tab:codes} summarizes the similarities and differences of
the five waveforms, and the codes used to produce them. More
details on the waveforms can be found in the following references: 
The {\tt BAM} waveform is from the highest-resolution D12 simulation
described in~\cite{Hannam:2007ik}.
The {\tt CCATIE} results have been obtained from the simulation
described in~\cite{Damour:2007vq}. 
%
The {\tt MayaKranc} simulation is the $e=0$ simulation described
in~\cite{Hinder:2007qu}. 
%
The {\tt SpEC} waveform corresponds to the waveform ``30c-1/N6''
described in Ref.~\cite{Scheel:2008rj}; the inspiral portion of this
waveform is more comprehensively discussed in
Ref.~\cite{Boyle:2007ft}, including a detailed error analysis.
The {\tt Hahndol} waveform comes from the highest-resolution (grid
spacing of $M/32$ at the finest level) evolution of the ``$d_i$ = 10.8 M''
data presented in \cite{Baker:2006kr}.
For all simulations the nominal Courant factor was 0.5, although for
 the {\tt BAM}, {\tt CCATIE} and {\tt MayaKranc} simulations the
 Courant factor was lowered on the two outermost mesh-refinement
 levels.

No new simulations were performed for this paper, although the {\tt
  MayaKranc} waveform results from an updated extrapolation procedure
as described in Section~\ref{sec:waves}. 

Table~\ref{tab:codes} also provides uncertainty estimates in the GW
phase and amplitude, quoted separately for the inspiral regime (up to
a frequency of roughly $M\omega = 0.2$), and the merger and ringdown
regime. It is important to bear in mind that each uncertainty estimate
applies only to the {\it waveform}, and not to the {\it code} used to
produce it. For example, a code that uses second-order-accurate finite
differcing may well produce waveforms more accurate than any presented
here, if run at sufficiently high resolution with sufficiently
accurate initial data, and if the gravitational waveforms were
extracted sufficiently far from the source. 

Note that the apparent accuracy of the phase and amplitude depend
strongly one how one chooses to align waveforms from different
simulations, and whether quantities are considered as functions of
time, phase or frequency. All of these choices are valid when
comparing results produced by evolving the same initial data with the
same evolution system, and varying only numerical resolution,
radiation extraction radii and outer boundary location, and one is
free to make the choice that gives the lowest error estimate. As such,
the methods used to estimate the phase and amplitude errors differ for
each waveform; more details can be found in some of the references
listed in Table~\ref{tab:codes}. 

Having said that, in the present study we are comparing results from
different codes, with different initial data and gauge conditions, and
the disagreements we see from different waveform alignment choices may
exaggerate, or hide, the ``real'' differences between the waveforms. 

We find that the least ambiguous method of comparison is to plot
quantities with respect to the frequency $M\omega$ of the $(\ell=2,m=2)$
mode of $\Psi_4$. This choice removes the need to apply a time and
phase shift when comparing the wave amplitude, and the freedom of a
constant phase shift in a phase comparison is straightforward to
interpret.

\begin{table*}
\caption{\label{tab:codes}
Summary of numerical codes. The initial separation is the coordinate
separation between the punctures (for moving-puncture codes) or
between the centers of the excision surfaces ({\tt SpEC}). 
The initial momenta
specified in the moving-puncture codes are $(p_t,p_r)/M$, where $p_t/M$ is
the tangential momentum and $p_r/M$ is the radial momentum. The {\tt
  SpEC} parameters are described in~\cite{Pfeiffer:2007yz}.
``Bulk FD order'' indicates the spatial finite difference order in the 
bulk of the computational domain (i.e., not including mesh-refinement
boundary zones). $h_{\rm min}$ is the spatial resolution on the finest
mesh-refinement level or domain. 
The wave extraction radii are given, and $r_{ex}
\rightarrow \infty$ indicates that the results were then extrapolated
to infinity. The references provide full details of the implementation
of the codes and the simulations that were used in this study.
For the {\tt CCATIE} result no numerical convergence results were published,
but based on the code specification and resolution for this run, a phase
accuracy between the {\tt Hahndol} and {\tt BAM}/{\tt MayaKranc} results can
be assumed. The amplitude errors quoted for {\tt CCATIE} and {\tt
  MayaKranc} were estimated for the present paper and were previously
unpublished. 
}
\begin{tabular}{||l|ll|c|c|c|c|c|c|c|c||}
\hline
Code         &  \multicolumn{2}{|c|}{Initial}           & Bulk  &
$h_{min}/M$ & Wave                     & eccentricity    & \multicolumn{2}{|c|}{Phase}         & \multicolumn{2}{|c||}{Amplitude}   \\  
                   &  \multicolumn{2}{|c|}{parameters} & FD  &
                   ($\times 10^{-3}$) & extraction             &                          & \multicolumn{2}{|c|}{uncertainty} & \multicolumn{2}{|c||}{uncertainty}     \\
                   &   \multicolumn{2}{|c|}{}
                   &  order  &                & radius                   &
                   & \multicolumn{2}{|c|}{(radians)}            &
                   \multicolumn{2}{|c||}{(percentage)}   \\
\hline
&\multicolumn{2}{|c|}{}&&&&
         & insp. & merger                          & insp.    & merger      \\
\hline
\multicolumn{11}{||l||}{Finite-difference moving-puncture codes} \\
\hline
{\tt BAM} \cite{Bruegmann:2006at,Husa2007a} & $D=12M$; &
$(0.085,-5.373\times10^{-4})$  & 6 & $19$ &  $90M$ & $e < 0.0016$  & $0.1$   &   1.0       & 4.0       &       6.0 \\

{\tt CCATIE} \cite{Pollney:2007ss} &  $D=11M$; & $(0.090,-7.094\times10^{-4})$ & 4 &  $20$ &$120 M$ & $e < 0.0016$ & & &  $2.0$  & $5.0$       \\

{\tt Hahndol} \cite{Imbiriba:2004tp,vanMeter:2006vi,Baker:2006ha}  &
$D=10.8M$;& $(0.0912, 0.0)$     & 4 &  $19$ & $60M$ & $e < 0.008$    & 2.4     &   5.0       &
10.0       & 10.0       \\
 
{\tt MayaKranc}  \cite{Vaishnav:2007nm} & $D=12M$;&
$(0.085,-5.343\times10^{-4})$  & 4 &  $15.5$ & $r_{ex} \rightarrow \infty$  & $e < 0.0016$  & 0.1 & 1.1  & 4.0  & 8.0   \\

\hline
\multicolumn{11}{||l||}{Pseudospectral excision code} \\
\hline

{\tt SpEC} \cite{Boyle:2007ft,Scheel:2008rj}     &
\multicolumn{2}{|p{5cm}|}{$D\!=\!14.436M$; \newline $r_{\rm
    exc}\!=\!0.41360M$, $M\Omega_0=0.016708$,\newline
  $v_r\!=\!-4.26\times 10^{-4}$, $f_r\!=\!0.939561$}   & n/a   &
$\sim 3$ & $r_{ex} \rightarrow \infty$ & $e < 5\times 10^{-5}$ &   0.006     &  0.02    &   0.1  &  0.3   \\
\hline
\end{tabular}
\end{table*}

\subsection{Extraction of gravitational waves}
\label{sec:waves}

In the numerical simulations presented here, the gravitational waves
are extracted using the Newman-Penrose Weyl tensor component
$\Psi_4$~\cite{Newman62a,Stewart:1990uf}, which 
at infinite separation from the source is related to  
the complex strain $\hc = h_+ - \mathrm{i} h_\times$ by~\cite{Teukolsky73},
\begin{equation}
  \hc = \lim_{r\rightarrow\infty} \int^t_0 \dt^\prime
  \int^{t^\prime}_0 \dt^{\prime\prime} \Psi_4.  \label{eq:psi4}
\end{equation}

Note that the amplitude of the gravitational-wave strain falls off as
$1/r$, where $r$ is the distance of the detector (or, in a numerical
code, the extraction sphere) to the source, and so we generally
consider $r \hc$ (and $r \Psi_4$), which in the weak-field region will
be independent of $r$. 

It is useful to discuss gravitational radiation fields in terms of
spherical harmonics of spin-weight $s=-2$, $Y^{s}_{\ell m}$, which represent
symmetric
tracefree 2-tensors on a sphere, and in this
paper we will only consider the dominant $\ell=2,\ m=\pm 2$ modes,
with basis functions 
\begin{eqnarray}
  Y^{-2}_{2-2} & \equiv & \sqrt{\frac{5}{64\pi}} \left(1 -\cos \theta \right)^2
       e^{-2 \mathrm{i}\varphi}, \nonumber \\
  Y^{-2}_{22} & \equiv & \sqrt{\frac{5}{64\pi}} \left( 1 +\cos \theta \right)^2
       e^{2 \mathrm{i}\varphi}\,,
\end{eqnarray}
i.e., we will consider the cases $\ell=2,m=\pm 2$ of the projections
\begin{equation}
\hc_{\ell m} \equiv \langle Y^{-2}_{\ell m}, \hc \rangle = 
        \int_0^{2\pi} \dphi \int_0^{\pi}
      \hc \, \overline{Y^{-2}_{\ell m}}\, \sin \theta\,\dtheta\,
      \label{eq: scalar_product},
\end{equation}
of the complex strain $\hc$ (bar denotes complex
conjugation).  In the nonspinning case considered here, we have equatorial
symmetry so that $\hc_{22} = \overline{ \hc_{2-2} }$, and
\begin{equation*}
  \hc(t) = \sqrt{\frac{5}{64\pi}} e^{2 \mathrm{i}\phi} \left(
    \left(1 + \cos \theta \right)^2       \hc_{22}(t)
    + \left(1 - \cos \theta \right)^2  \bar \hc_{22}(t) \right). 
\label{eq:NRh}
\end{equation*}

The coordinate radius at which the waves were extracted from the 
numerical solution is given for each code in Table~\ref{tab:codes}. 
For the {\tt MayaKranc} and {\tt SpEC} codes, the waves were extracted
at several radii, and then extrapolated to $r_{ex} \rightarrow \infty$, to
give a more accurate estimate of the wave that would be measured 
by a distant GW detector. The extrapolation procedure involves 
aligning the waveforms with respect to some definition of retarded
time \cite{Boyle:2007ft}, and then treating the error due to extraction radius as a
polynomial in powers of $1/r_{ex}$ \cite{Hannam:2007ik,Boyle:2007ft}. 
Different polynomial fits were performed for the inspiral and merger for 
the {\tt MayaKranc} waveform, and the specific extrapolation procedure 
used for the  {\tt SpEC} waveform is given in \cite{Boyle:2007ft,Scheel:2008rj}.
Waves were extracted from the {\tt CCATIE} simulation using the Zerilli-Moncrief
procedure (see~\cite{Nagar:2005ea} for a review), from which $\Psi_4$ 
can be readily derived.   

The direct waveform comparisons in Section~\ref{sec:phase} deal with
$r\Psi_4$. The data-analysis comparisons in Section~\ref{sec:match}
are based on the strain, $r\hc$. To produce the strain from $\Psi_4$
one needs ``merely'' to integrate twice with respect to time, as in
Eq.~(\ref{eq:psi4}), and choose appropriate integration 
constants. However, this procedure is not as trivial as it at first
appears. One might naively assume that integration constants could be
chosen on simple physical grounds, for example that the strain rings
down to zero after the black holes have merged, and that it oscillate
around zero at all times. Such requirements have been found to work
adequately in some cases for the $(\ell=2,m=2)$ mode, but even in the
best cases unusual artifacts remain, and these become more pronounced
when one considers higher modes; see \cite{Berti:2007fi} for some
examples. One reason for these difficulties is that the waveforms
contain small numerical errors and gauge effects, which become greatly
exaggerated when integrated over the entire duration of the waveform
--- and to calculate the strain we must perform such an integration
twice. 

A tempting alternative is to work only in the Fourier domain. Start
with the numerically generated $\Psi_4(t)$, calculate the Fourier
transform, $\tilde{\Psi}_4(f)$, and then it is trivial to perform two
time integrations to obtain the Fourier transform of the strain, \beq
\tilde{\hc}(f) = - \frac{\tilde{\Psi}_4(f)}{4 \pi^2 f^2}.
\eeq The integration constants have been ignored in this procedure,
or, rather, they have been implicitly set to zero. If we now perform
an inverse Fourier transform to calculate $\hc(t)$, we will recover
similar artifacts to those we would have seen if we had performed two
time integrations of $\Psi_4(t)$. 

However, in this paper we use this very method to calculate
$\tilde{\hc}(f)$ to use in our match calculations. Our justification is
that for a selection of waveforms we have independently calculated
$\hc(t)$ by a number of different methods (with varying levels of
success in removing numerical and gauge artifacts in the final
strain), and have then used the Fourier transform of this quantity in
match calculations, and obtained very similar results; we will discuss
the impact of the small differences that we see in Section~\ref{sec:match}.  
Our conclusion is that the choice of integration constants, and
modifications that ``clean'' the waveform of non-physical artifacts,
although they may lead to serious differences in the time-domain waveform,
do not significantly affect the match calculation for the $(\ell=2,m=2)$
mode.

\section{Direct comparison of phase and amplitude}
\label{sec:phase}

We now compare the waveforms produced by the five codes. For the
purposes of gravitational-wave detection, the most meaningful
comparison will include the noise spectrum of the detector. We will
make comparisons relevant to detection and parameter estimation 
in Section~\ref{sec:match}. In this section we
directly compare the numerical waveforms in a manner that is independent
of any particular detector.  The quantities we will compare are the amplitude
$A(t)$ and the phase $\phi(t)$ of the $(\ell=2,m=2)$ mode of
$r\Psi_4$, which are defined by
\begin{equation}
r \Psi_{4,22}(t) = A(t) e^{-i \phi(t)}.
\end{equation} The GW frequency for the $(\ell=2,m=2)$ mode is given
by $\omega(t) = \dot{\phi}(t)$. 

The amplitude and phase are the two pieces of
raw output from the computer code that define the waveform, so they allow the
most direct comparison between results from different codes. More
generally, an amplitude/phase comparison allows us to
quantify waveform differences independent of any detector --- if two
waveforms accumulate one cycle of phase disagreement during the last
ten cycles before they reach the peak amplitude, that is a difference
that will exist no matter which detector they pass through. 

On the other hand, there are a number of ambiguities in an 
amplitude/phase comparison, which we will describe as we proceed. For
the purpose of gravitational-wave detection, the detector mismatch and 
SNR are more meaningful quantities to compare. We can summarize the
situation as follows: a direct comparison of amplitude and phase is
most useful to the numerical relativist, while the mismatch and SNR are
most useful to the data analyst.

\subsection{Phase} 

In comparing the phases of two waveforms from different simulations,
$\phi_1(t_1)$ and $\phi_2(t_2)$, 
we cannot simply calculate $\phi_1(t_1) -
\phi_2(t_2)$, because the time coordinates $t_1$ and $t_2$ may not be
the same. The two simulations may have been started at different
points along the binary inspiral, meaning that $t_1=0$ does not label
the same event as $t_2=0$. More simply: although the waveforms may be
identical, one will reach a detector later than the other. 

To make a comparison we first have to decide on an event at which the two 
waveforms should agree, and to then apply a relative time shift  
and phase shift so that the chosen event occurs at the same time 
and phase for each waveform. For example, if we were to align the phases at the time
when the waveform amplitude reaches a maximum, then we would first 
determine the times $T_1$ and $T_2$ when each waveform's amplitude 
reaches its maximum, and then study the quantity \beq
\Delta \phi(t) = \phi_1 (t + T_1) - \phi_2(t + T_2) + \phi(T_2) - \phi(T_1),
\eeq where by construction the amplitude maxima now occur at $t=0$ and
$\Delta \phi(0) = 0$.

The problem with this procedure is that $\Delta \phi(t)$ is extremely sensitive
to the accuracy with which $T_1$ and $T_2$ were determined, particularly
around the merger, when the GW frequency increases rapidly. One solution 
is to make a further small time shift, until the overall phase disagreement between 
the two waveforms has been minimized. Such a suggestion has been used in the 
past in matching NR and PN waveforms \cite{Ajith:2007qp,Boyle:2009dg},
and in NR-PN 
comparisons \cite{Boyle:2008ge}. This, however, is an approach designed not
to determine the differences between two waveforms, but to minimize them. 
Another option, which avoids the time-shift ambiguity 
altogether, is to compare the phases as a function of GW frequency $\phi(M\omega)$; 
this procedure was used in \cite{Baker:2006ha}, and we will use it here.

\begin{figure}[t]
\centering
\includegraphics[width=85mm]{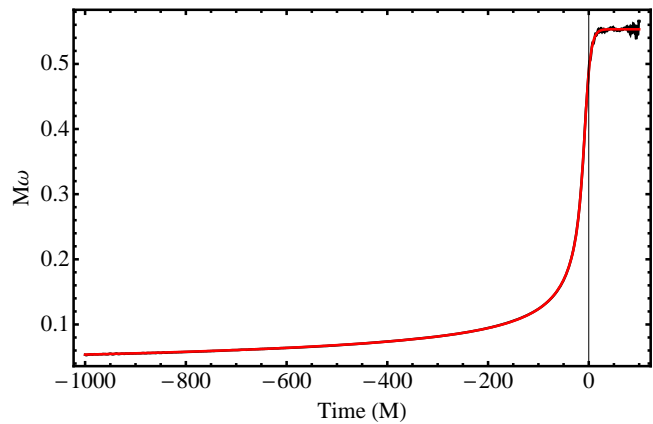}
\caption{The GW frequency as calculated from the raw {\tt BAM} data,
  and as given by the fitting procedure described in the text. The two
lines are indistinguishable if viewed in black-and-white.}
\label{fig:Omega}
\end{figure}

\begin{figure}[t]
\centering
\includegraphics[width=85mm,bb=0 10 445 305]{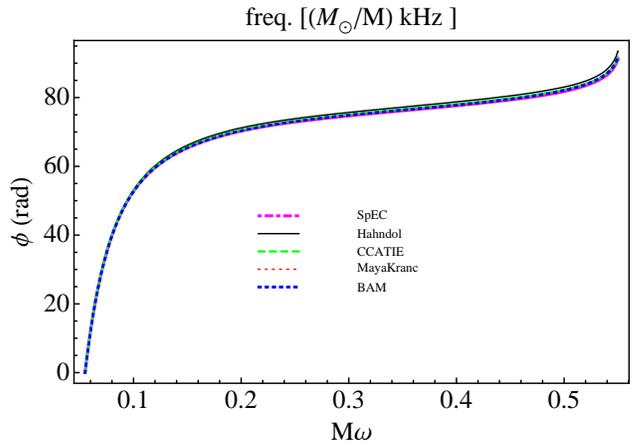}
\caption{GW phase $\phi$ as a function of frequency $M\omega$, for the
five codes. The frequency is given both in terms
 of the dimensionless orbital frequency $M\omega$, and the frequency in
 kHz scaled with respect to the total mass of the binary in solar masses.}
\label{fig:Phase}
\end{figure}

The GW frequency as read from the numerical data is too noisy at early and 
late times to allow a clean direct parametric plotting of phase vs frequency.
We instead fit the
frequency to a combination of the TaylorT3 PN frequency formula
\cite{Buonanno:2006ui} and a modification of the frequency ansatz
introduced in \cite{Baker:2008mj}. Specifically, the TaylorT3
expression for the orbital frequency of the binary during inspiral is
given up to 3.5PN order by \cite{Blanchet:2004ek,Buonanno:2006ui} 
\begin{eqnarray}
\label{eqn:T3}
& &\Omega_{PN}(\tau) \nonumber\\
& & =  \frac{1}{8} \tau^{-3/8} \left[ 
  1 
+ \left( \frac{743}{2688} + \frac{11}{32} \nu \right) \tau^{-1/4} 
- \frac{3}{10} \pi  \tau^{-3/8} \right. \nonumber  \\
& &\left.  + \left( \frac{1855099}{14450688} + \frac{56975}{258048} \nu +
  \frac{371}{2048} \nu^2 \right) \tau^{-1/2} \right. \nonumber \\
& & \left. + \left( - \frac{7729}{21504} + \frac{13}{256} \nu \right)
  \pi \tau^{-5/8}  \right. \nonumber  \\
& & \left. + \left( - \frac{720817631400877}{288412611379200} + \frac{53}{200}\pi^2 
 + \frac{107}{280} \gamma \right. \right. \nonumber \\ 
& & \left. \left. - \frac{107}{2240} \ln \left(
       \frac{\tau}{256} \right)  
+ \left( \frac{25302017977}{4161798144} - \frac{451}{2048} \pi^2
\right) \nu \right. \right. \nonumber \\
& & \left. \left. - \frac{30913}{1835008} \nu^2 + \frac{235925}{1769472} \nu^3\right) \tau^{-3/4}
  + a(\nu) \tau^{-7/8}  \right],
\end{eqnarray} where we have given the last (3.5PN) term an arbitrary
coefficient, $a(\nu)$, where $\nu$ is the symmetric mass ratio 
$\nu = m_1 m_2 / M^2$. This term is known in PN theory, but we will instead
fit it to our numerical data, given our modified definition of the variable $\tau$,
which we will now discuss. The definition of $\tau$ in standard PN theory is
\beq
\tau = \frac{\nu (t_c - t)}{5M},
\eeq
where $t_c$ is a PN estimate of the ``coalescence time''. 
The expression (\ref{eqn:T3}) diverges when $\tau = 0$, so in order to produce a
formula which can be fit through our data, we use instead
\beq
\tau^2 = \frac{\nu^2 (t_c - t)^2}{25 M^2} + 1,
\eeq
and we now treat $t_c$ as a parameter to fit to the data, as in
\cite{Buonanno:2006ui}. With our new definition of $\tau$, the expression
(\ref{eqn:T3}) becomes inaccurate near $\tau=0$ (which is anyway true
for any post-Newtonian expression near merger), but does not
diverge. To model the ringdown phase, we modify the ansatz suggested
in \cite{Baker:2008mj}, and write the full frequency
as \begin{eqnarray}
\Omega(t) & = & \Omega_{PN}(\tau) + \nonumber \\ 
&& \left(\Omega_f - \Omega_{PN}(\tau)\right) 
\left( \frac{1 + \tanh[\ln\sqrt{\kappa} - (t - t_0)/b]}{2} \right)^\kappa. \nonumber \\
\label{eqn:omfit}
\end{eqnarray} The constants $\{t_c,t_0,\kappa,a,b,\Omega_f\}$ are
parameters that are determined to produce the best fit to the
numerical data. The constant $\Omega_f$ corresponds to a fit of the
ringdown frequency. The frequency as a function of time is shown for
the {\tt BAM} code in Fig.~\ref{fig:Omega}, as calculated from the raw
numerical data, and as given by the fitting procedure we have just
described; the GW frequency is related to the orbital frequency by a
factor of two. The GW phase as a function of frequency for each of the
five waveforms is shown in Fig.~\ref{fig:Phase}.

\begin{figure*}[t]
\centering
\includegraphics[width=84mm,bb=0 10 430 310]{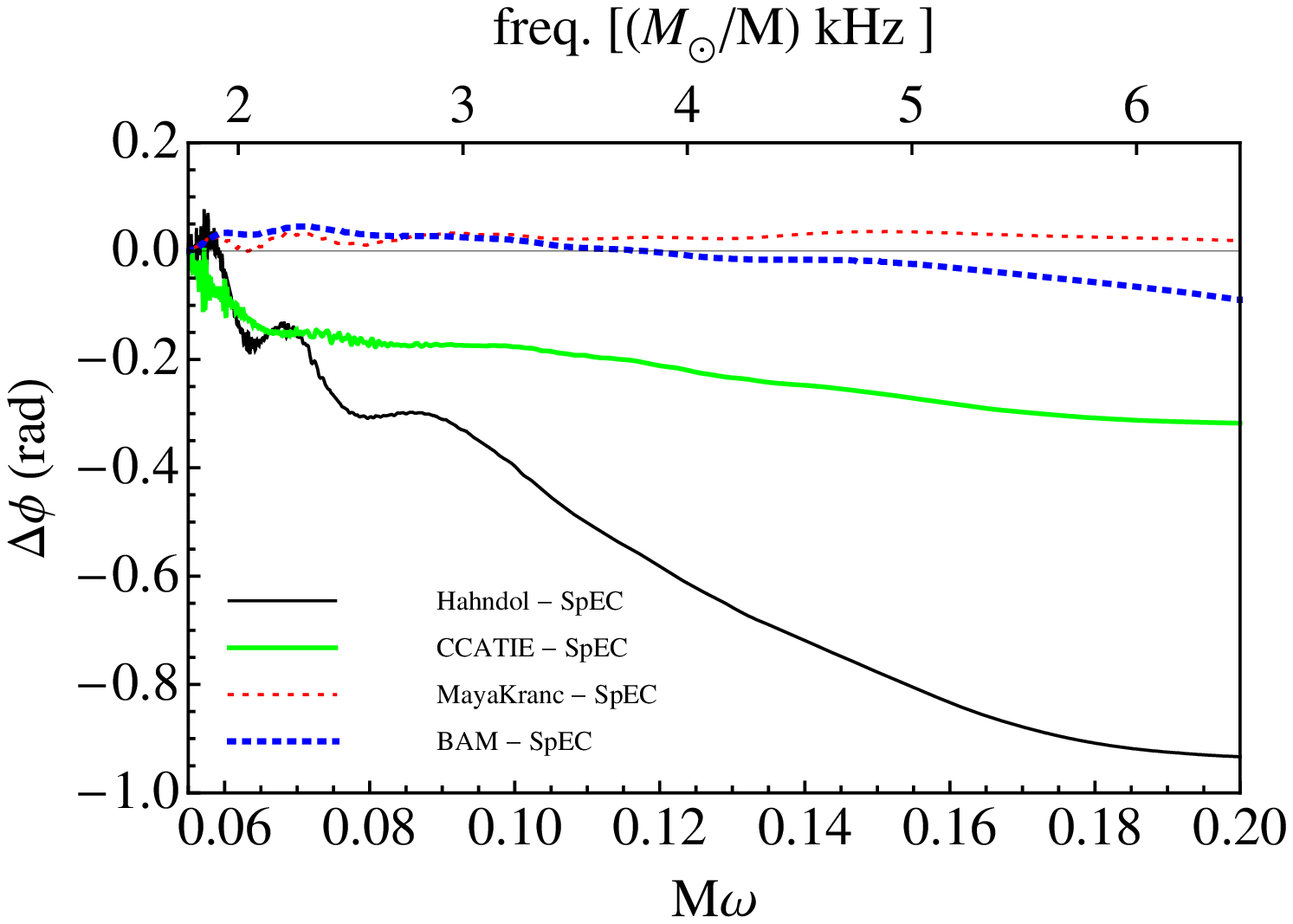}
$\qquad$
\includegraphics[width=84mm,bb=0 10 430 305]{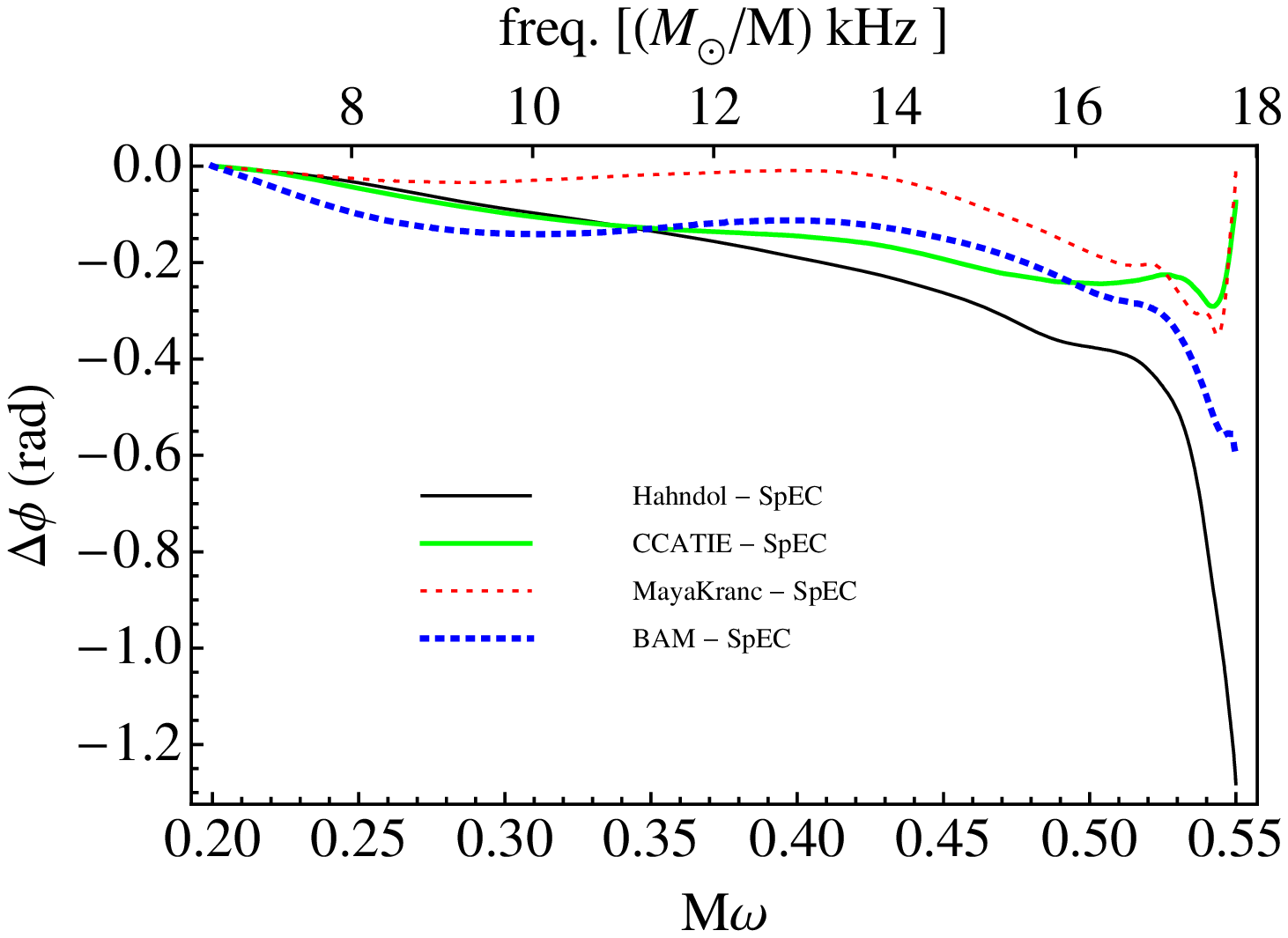}
\caption{Phase comparison. The left panel shows the phase comparison 
between the {\tt SpEC} waveform and the others during inspiral, 
from $M\omega\! =\! 0.055$ up to $M\omega\! =\! 0.2$, which is about 
one orbit before merger. The corresponding uncertainty estimates are 0.1~rad ({\tt BAM} and {\tt MayaKranc}) and 2.4~rad ({\tt
  Hahndol}). The right panel shows the phase comparison 
during merger and ringdown, 
from $M\omega\! =\! 0.2$ up to $M\omega\! =\! 0.55$. The uncertainty estimates
during merger are, in radians, 1.0 ({\tt BAM}), 1.1 ({\tt
  MayaKranc}), and 5.0 ({\tt Hahndol}). In both panels a 
phase shift was applied so that the phases agreed at the lowest frequency
shown.}
\label{fig:PhaseComparison}
\end{figure*}

Figure~\ref{fig:PhaseComparison}  compares the phase of each waveform 
with that from the {\tt SpEC} code. The GW frequency $1000M$ before merger,
where our waveforms nominally begin, is close to $M\omega = 0.055$, and this
is the frequency at which our comparison begins. After merger, the merged 
black hole rings down to the Kerr solution, and the GWs are emitted at the 
ringdown frequency, which is close to $M\omega = 0.55$; this is where we 
end our comparison. (The precise ringdown frequency for the equal-mass,
nonspinning, zero-eccentricity configuration is 
$M\omega = 0.5535$ \cite{Scheel:2008rj}.)

In the left panel of Figure~\ref{fig:PhaseComparison} we show the phase
disagreement during inspiral, ending at $M\omega = 0.2$, which is reached 
about half an orbit before merger. A phase shift is applied so that the phases all agree at 
$M\omega = 0.055$. We see that the accumulated phase disagreement is 
below 0.3\,radians for all codes except {\tt Hahndol}, for which the larger
eccentricity and lower resolution lead to larger dephasing against the 
{\tt SpEC} results. The
behaviour of the three other waveforms is roughly consistent with the
numerical methods used to produce them: the {\tt BAM} waveform was produced
with the highest-order spatial finite-differencing (sixth-order), and while 
fourth-order spatial finite-differencing was used to produce both the
{\tt CCATIE} and {\tt MayaKranc} results, the {\tt MayaKranc} simulation 
was performed at slightly higher resolution, and the results were further
extrapolated with respect to radiation extraction radius. 

The most important point is that the results of each code agree within their
respective uncertainty estimates. 

The right panel of Figure~\ref{fig:PhaseComparison} shows the accumulated
phase disagreement during the last orbit, merger and ringdown. The phases
are shifted to agree at the lowest frequency shown in the figure,
$M\omega = 0.2$, so that we can see how the phase disagreement behaves
during the merger regime only. Note that the waveform from the {\tt Hahndol}
simulation becomes very noisy late in the ringdown, which accounts for
the poor behaviour above $M\omega \approx 0.52$. 
Note also that while the merger and ringdown plot 
sweeps through roughly twice the range of frequencies as the inspiral plot, the
length of time covered during the inspiral (about $900M$) is much {\it greater} 
than that during the merger (about $180M$). In this sense the phase 
disagreement grows more quickly during merger. The phase disagreements of
the different waveforms are again consistent
with uncertainty estimates, but are larger for merger and ringdown than 
they were during inspiral.

The two panels of Fig.~\ref{fig:PhaseComparison} were designed to show
separately the phase difference accumulated during inspiral, or during
merger/ring-down.  When considering the phase as a function of
frequency (as done in Fig.~\ref{fig:PhaseComparison}), the only
freedom is an overall additive constant to the phase.  Thus, the total
accumulated phase difference during inspiral {\em and} merger/ringdown
can be obtained by vertically offsetting the curves in the right panel
of Fig.~\ref{fig:PhaseComparison}, so that the phase-differences at
$M\omega=0.2$ agree in both panels.  For instance, the total
accumulated phase-difference between {\tt BAM} and {\tt SpEC} at 
$M\omega = 0.52$ would be
the sum of 0.1~rad (from the left panel of
Fig.~\ref{fig:PhaseComparison}), and 0.28~rad (from the right panel),
i.e., 0.38~rad.  For the other codes, one finds at $M\omega = 0.52$ the 
following total accumulated phase-differences relative to {\tt SpEC}: 
{\tt Hahndol} 1.36~rad, {\tt CCATIE} 0.55~rad, and {\tt MayaKranc} 0.18~rad.
The reader may choose to calculate the total accumulated phase
disagreement at any frequency, although one should bear in mind that
beyond $M\omega = 0.52$ the curves in Fig.~\ref{fig:PhaseComparison} 
are less reliable, due to errors in the curve fit Eqn~(\ref{eqn:omfit}) through
noisy numerical data.

\subsection{Amplitude}

\begin{figure}[b]
\centering
\includegraphics[width=85mm,bb=5 12 420 305]{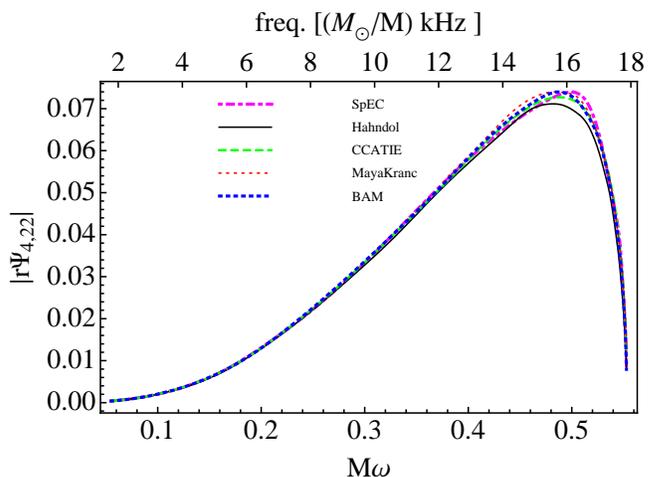}
\caption{The amplitude as a function of GW frequency, $A(\omega) =
  |r\Psi_{4,22}|$ for the five codes.}
\label{fig:Amplitudes}
\end{figure}

\begin{figure*}[t]
\centering
\includegraphics[width=85mm,bb=0 8 440 305]{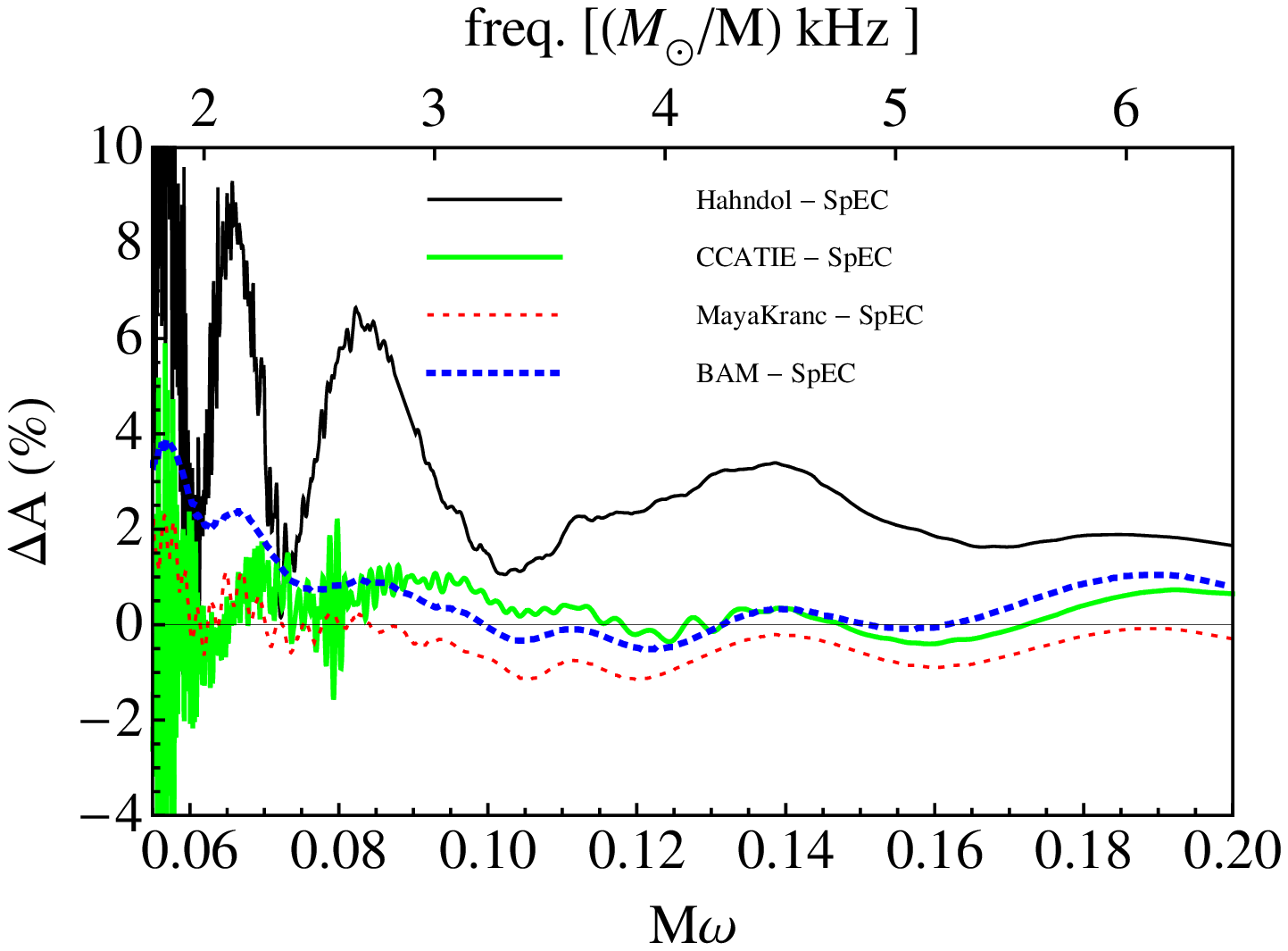}
$\quad$
\includegraphics[width=85mm,bb=0 8 440 305]{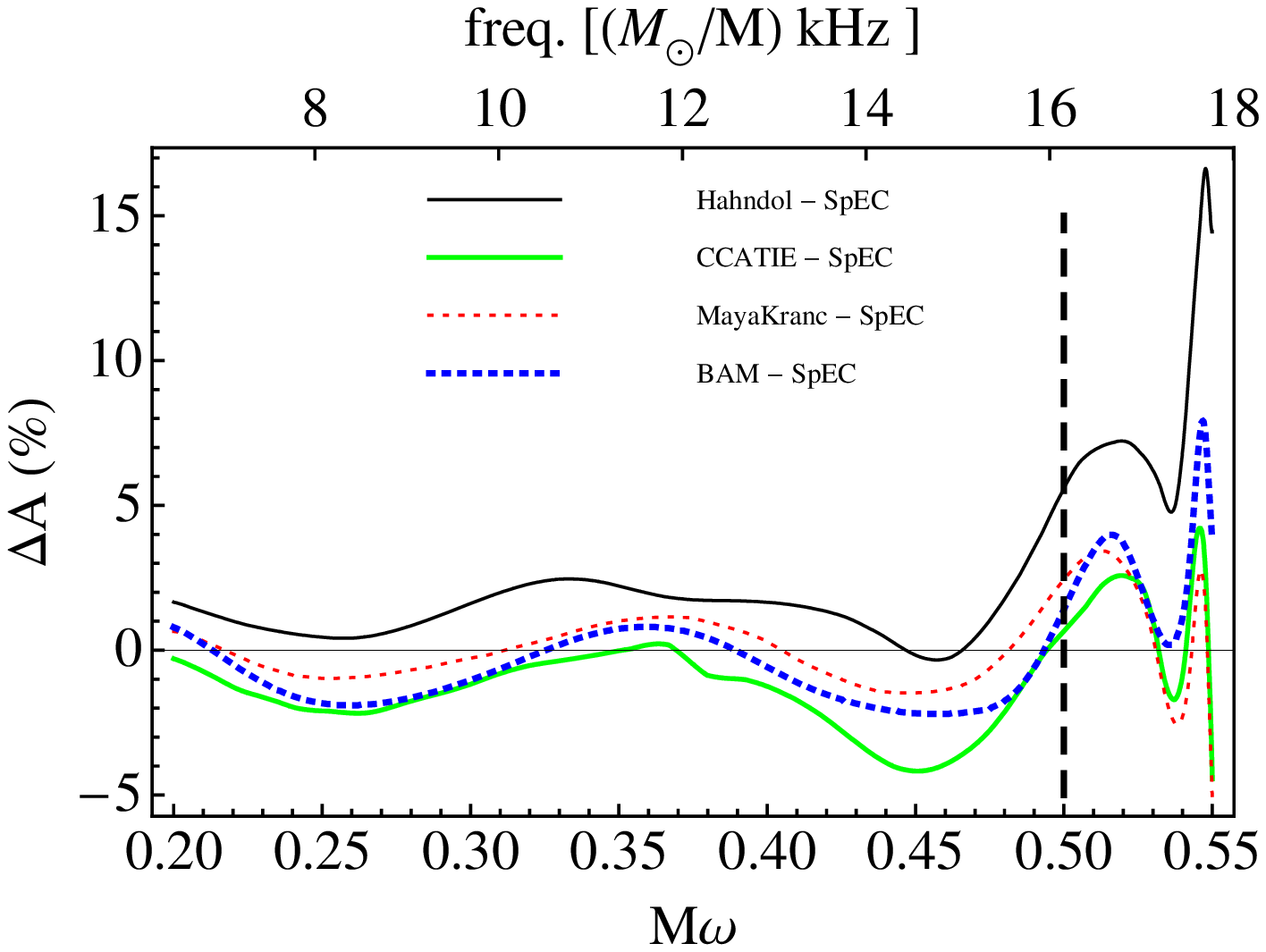}
\caption{Comparison of the amplitude as a function of GW frequency,
  $A(\omega)$. The left panel shows the percentage disagreement
  during inspiral (up to $M\omega = 0.2$). The corresponding amplitude
  uncertainties are 2\% ({\tt CCATIE}), 4\% ({\tt BAM} and {\tt
    MayaKranc}), and 10\% ({\tt Hahndol}). 
  The right panel shows
  the same quantity during merger and ringdown, for which the
  uncertainties are 5\% ({\tt CCATIE}), 6\% ({\tt BAM}), 8\% ({\tt
    MayaKranc}), and 10\% ({\tt Hahndol}).  
  The vertical dashed
  line indicates the approximate location of the amplitude maximum,
  $M\omega = 0.5$. 
}
\label{fig:AmpComparison}
\end{figure*}

In comparing the GW amplitude between codes, we once again consider
the amplitude as a function of frequency, $A(M\omega)$, which is shown
for the five codes in Fig.~\ref{fig:Amplitudes}. The amplitude comparison 
is shown in Fig.~\ref{fig:AmpComparison}. Once again the comparison
during inspiral, shown in the left panel, covers the frequency range
$M\omega \in [0.055,0.2]$, and the comparison during merger and
ringdown, shown in the right panel, covers the frequency range
$M\omega \in [0.2,0.55]$. The oscillations in the right panel of 
Fig.~\ref{fig:AmpComparison} are probably due to small errors
from gauge effects that are exaggerated in this plot by the rapid change
in the GW frequency near merger. Recall that the {\tt
  Hahndol} waveform becomes unreliable at about $M\omega \approx 0.52$. 

Note once again that the agreement is within the estimated 
uncertainties of the waveforms.

The conclusion of our direct comparison of the GW phase and amplitude
is that all five codes are consistent within their stated error
bars. This provides an important consistency check on the numerical 
accuracy and validity of each waveform. Not only that, it provides us
with an upper limit on the variations in the waveforms due to different
choices of initial data (in these cases puncture data versus
quasi-equilibrium conformal-thin-sandwich excision data), and different
gauge choices. The latter can lead to noticeable differences in the 
amplitude and phase of the waveform extracted from the simulation (see,
for example, the discussions in 
\cite{Bruegmann:2006at,Scheel:2008rj,Lehner:2007ip}).
 With suitable gauge choices these differences
should decrease as the waves are extracted successively further from the 
source, and indeed it is usually possible to perform some procedure to
extrapolate the waveform to estimate the result that would be measured
infinitely far away
\cite{Hannam:2007ik,Boyle:2007ft,Hinder:2008kv,Scheel:2008rj}.  
These are delicate procedures, and one may 
still worry that the different gauge choices between codes will lead to 
large differences in the final waveforms. In this section we have shown that, 
if such differences exist, they are small and within the error bars of each 
simulation. 

The results so far provide information that
allow numerical relativists to quantify the accuracy and consistency 
of their results. In the next section we will make comparisons relevant
to data analysis and GW astronomy.

\section{Detection}
\label{sec:match}

A more meaningful comparison from the point of view of GW {\it
  detection} is the best match (and mismatch) between 
waveforms~\cite{Owen_B:96}.

The match is usually calculated in the frequency
domain. Consider two time series $x(t)$ and $y(t)$, which will be the
two waveforms we wish to compare. The Fourier transform is given by   
\begin{equation}
   \label{eq:FT}
   \tilde{x}(f) = \int_{-\infty}^\infty x(t)e^{2\pi if t}dt\, .
 \end{equation} 
(In LSC applications the opposite sign convention is used for the phase
in the Fourier transform definition, but the choice of sign does not affect 
the results here.)
In practice we calculate a discrete Fourier transform on the
numerical data. We calculate the time when the wave amplitude reaches
its maximum, $t_{max}$, and then truncate the waveform $1000M$ before
this time, and $80M$ after. The resulting truncated waveform is then
resampled every $0.1M$, to give a data set with 10,800 points. We
then take a discrete Fourier transform of each such data set, and
retain only the half of the data set that covers positive
frequencies. We also verified that our results did not change 
significantly when the sampling rate was varied. 

We can define an inner product between $\tilde{x}(f)$ and $\tilde{y}(f)$
weighted with the noise spectrum of the detector, $S_n(f)$~\cite{Cutler94},
 \beq
   \label{eq:scalar_prod}
 \langle x|y\rangle := 4 \, {\rm Re} \left[ \int_{f_{\rm
         min}}^{f_{\rm max}} 
     \frac{ \tilde x(f) 
     \tilde y^\star (f)}{S_n(f)} \, df \right] \, .
 \eeq
In the same way we define a norm of a waveform $\tilde{x}(f)$ by
 $|x| = \sqrt{\langle x | x \rangle}$.
 
The signal-to-noise ratio (SNR) is defined with respect to this waveform norm. 
Recall that
throughout this paper we have been dealing with $r \hc$ and $r
\Psi_4$, where $r$ is the distance of the detector (or numerical wave
extraction) from the source, and one should remember to use the real
strain $\hc$ in the definition of the SNR. For clarity, let us define
$\bar{\hc} = r \hc$, as calculated from the numerical code, and then
the SNR is given by \begin{equation}
\label{eqn:SNR}
\rho = \frac{\sqrt{\langle \bar{h}|\bar{h}\rangle}}{R},
\end{equation} where $R$ is the distance of the source from the
detector, usually in units of Mpc. 
 
The best match~\cite{Owen_B:96} is defined as the inner product $\langle x | 
y \rangle$ normalized by the norms of each waveform, and
maximized over relative time and phase shifts ($\tau$ and $\Phi$) 
between the two waveforms:
 \begin{equation}
 {\cal M} = \max_{\tau,\Phi} \frac{\langle x | y \rangle }{\sqrt{
     \langle x|x\rangle \langle y|y\rangle }}. 
 \end{equation} 
We can view this procedure as adjusting the waveforms
 with respect to their time of arrival, and their initial phase, such
 that we achieve the best agreement. 

\begin{figure*}[tp]
\centering
\includegraphics[width=80mm]{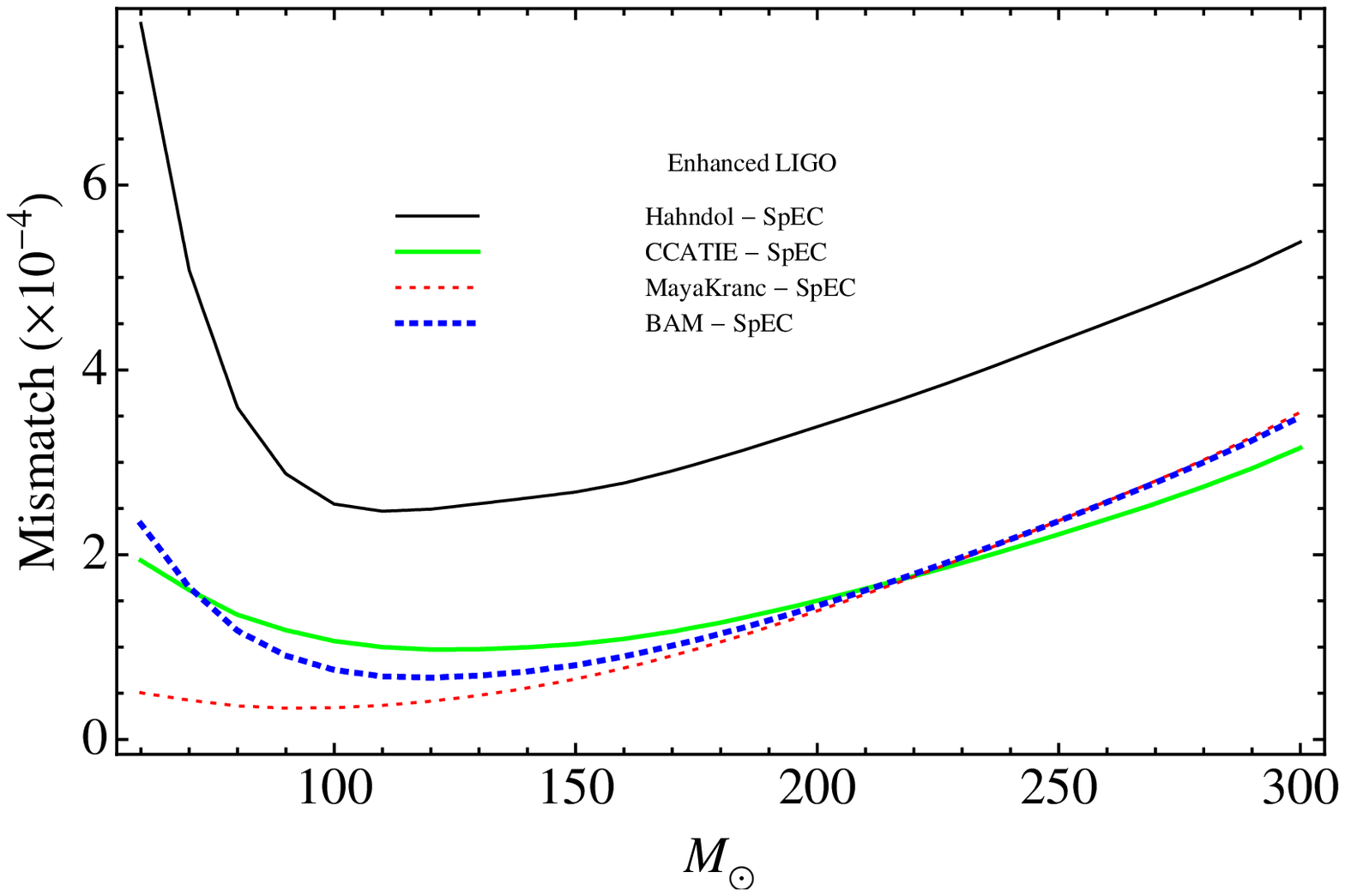}$\quad$
\includegraphics[width=80mm]{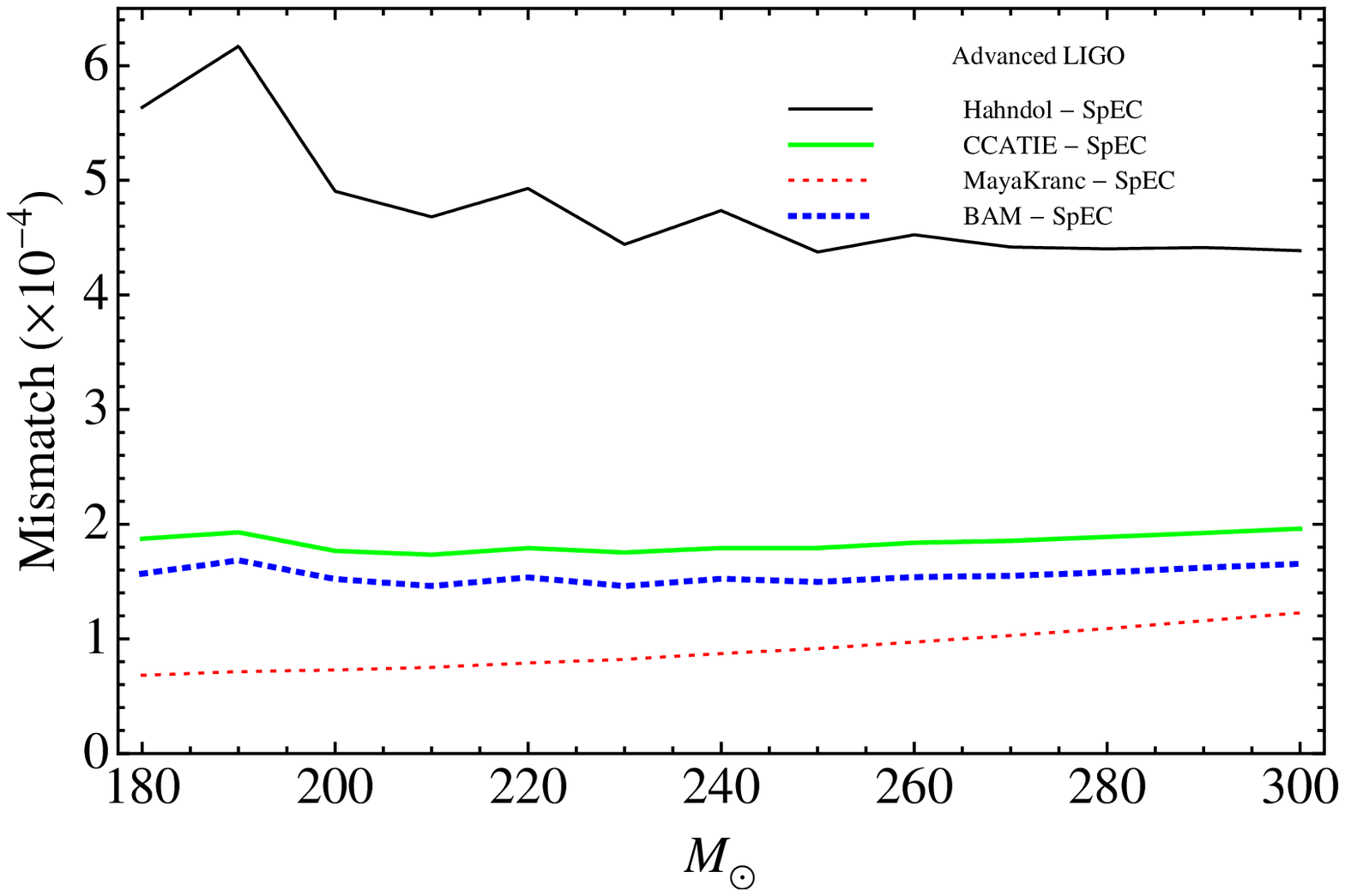}
\includegraphics[width=80mm]{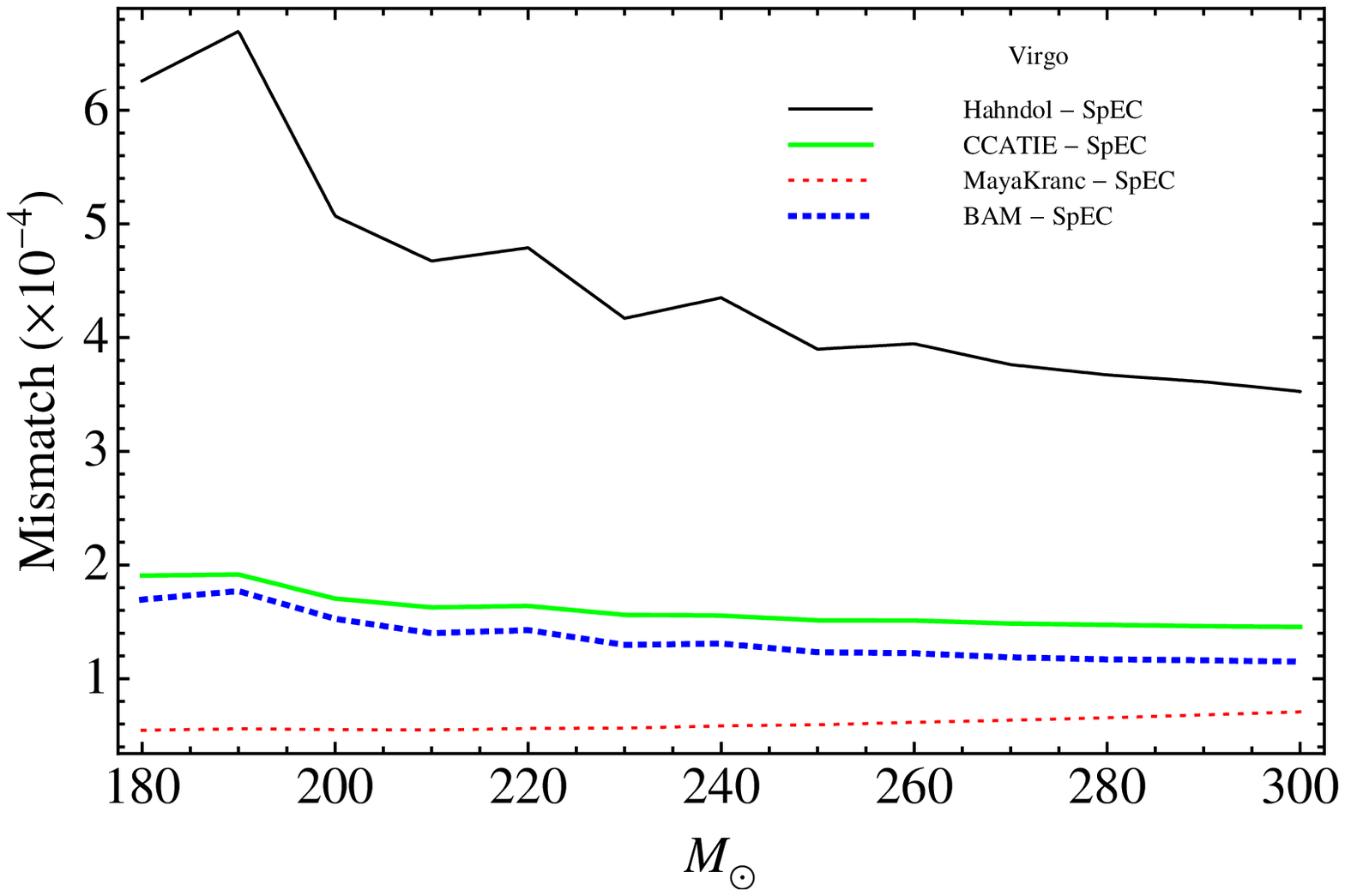}$\quad$
\includegraphics[width=80mm]{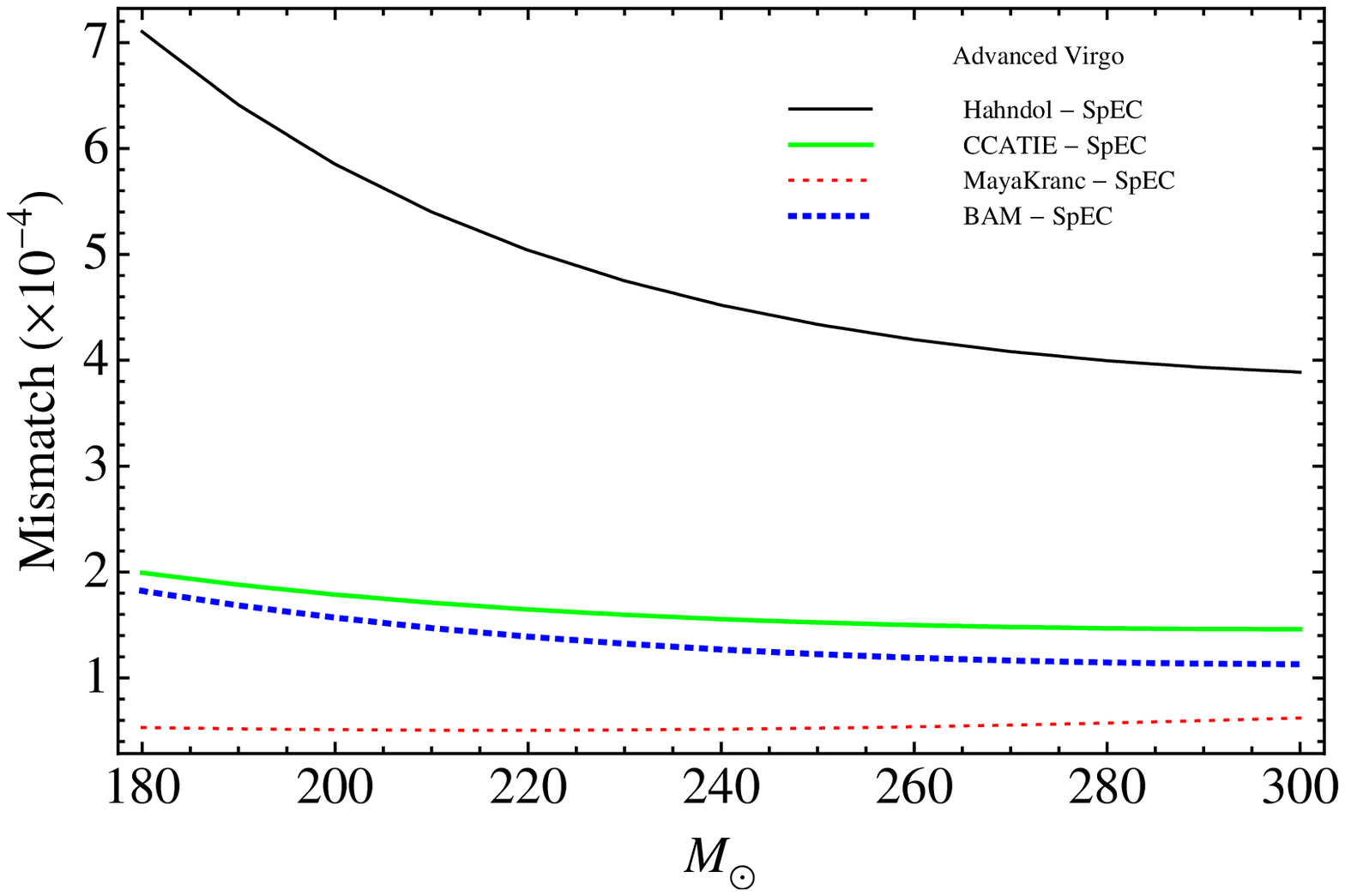}
\caption{The mismatch between the {\tt SpEC} waveform and each of the
  other codes (the line colors match those in previous plots). The
  three plots show the results for the Enhanced LIGO, Advanced LIGO,
  Virgo and Advanced Virgo noise curves. The lower end of the mass range was chosen
  such that the entire numerical waveform was included in the
  detector's frequency band.}
\label{fig:matches}
\end{figure*}

One convenient way \cite{Damour_T:98} to calculate the best match with 
respect to the phase shift is to first normalize 
each polarization of the two waveforms as $e_{1+,\times} =
\tilde{x}_{+,\times} / |x_{+,\times}|$ and $e_{2+,\times} = 
\tilde{y}_{+,\times} / | y_{+,\times} |$, and to
define
\begin{eqnarray*} 
A & = & \langle e_{1+}|e_{2+}\rangle^2 + \langle
e_{1+}|e_{2\times}\rangle^2, \\ 
B & = & \langle e_{1\times}|e_{2+}\rangle^2 + \langle e_{1\times}|e_{2
  \times}\rangle^2, \\ 
C & = & \langle e_{1+}|e_{2+}\rangle\langle e_{1\times}|e_{2+}\rangle +
\langle e_{1+}|e_{2\times}\rangle\langle e_{1\times}|e_{2\times}\rangle.
\end{eqnarray*}
In general one should also orthornomalize the two
waveforms, but in this work we consider only optimally oriented
binaries, and so this is not necessary. The best match is then given
by \cite{Damour_T:98} \begin{equation}
{\cal M} = \max_{\tau} \left[ \frac{A+B}{2} - \left[\left(
      \frac{A-B}{2} \right)^2 + C^2 \right]^{\frac{1}{2}}
\right]^{\frac{1}{2}}. 
\end{equation}

For the waveforms we consider here the match is very
close to unity, and it makes more sense to quote the mismatch, defined
as $1 - {\cal M}$. 

In evaluating 
(\ref{eq:scalar_prod}), we must choose
$f_{\rm min}$ and $f_{\rm max}$. Ideally these would be $(0,\infty)$,
but in practice they are based on the range of frequencies for which the 
Fourier transform is reliable (very high and very low frequencies
contain unphysical artifacts due to numerical errors and the sampling
rate of the data), and the relevant frequency window of a given
detector. The frequency range for the Enhanced LIGO detector is chosen
as 30~Hz up to 2~kHz, and for the Advanced LIGO,  Virgo and Advanced 
Virgo
detectors it is from 10~Hz up to 2~kHz. The acceptable frequency range
for the Fourier transforms of the numerical data is from $fM = 0.001$
up to $fM = 1$. The actual integrals are performed over the
intersection of the detector and waveform frequency ranges. 

The time shift $\tau$ deserves some discussion. From our
experience with phase comparisons in Section~\ref{sec:phase} we know that
in general we must time-shift two waveforms with respect to each other
in order to realistically estimate their agreement. It is natural to
apply that time shift to one of the waveforms in the time domain, but
if we have the entire frequency-domain representation of the signal,
we can also calculate the effect of a time shift on the match
in the frequency domain. In our case we do {\it  not} have the 
entire frequency-domain waveform: the numerical waveform was truncated
$1000M$ before merger, and $80M$ after merger, and as such the
calculated Fourier power is incorrect outside a certain range of
frequencies. Put another way, a time shift in the time domain would
result in using in our analysis a different $1080M$-long portion of
one of the waveforms, and there is no way that the analysis of the Fourier
transform of one $1080M$-long segment of the waveform can
capture the effect of choosing a different $1080M$-long segment. 
As such we apply a time shift to one of the waveforms {\it before}
calculating its Fourier transform, and maximise the match with respect
to this time shift. 

Figure~\ref{fig:matches} shows the minimum mismatch between the {\tt
  SpEC} waveform and each of the other four, for the Enhanced LIGO,
Advanced LIGO, Virgo and Advanced Virgo noise curves. A  mass range of 
$60-300\,M_{\odot}$ was used for the Enhanced LIGO detector, and 
$180-300\,M_{\odot}$ for the others. The lower mass was dictated by
the desire that the waveform begin below the low-frequency cut-off of 
the detector. At yet lower masses we would need to use longer
waveforms, i.e., waveforms that extend to lower frequencies. We can
see from these results that the mismatches are excellent: below
$10^{-3}$ in all cases, and for all except the {\tt Hahndol} waveform,
the mismatch is below $4 \times 10^{-4}$. The performance of each 
waveform is consistent with the expected accuracy of each code, and
with the results presented in Section~\ref{sec:phase}. Recall also that 
the mismatch calculation usually involves 
adjusting the mass associated with one of the 
waveforms, in order to improve the result. If such a minimisation
(or, in terms of the match, a maximisation) were performed here, the
mismatches would improve further.  

The mismatch is in general very sensitive to phase differences, and this
may explain the worse mismatch of the {\tt Hahndol} waveform, which shows 
large variations in the phase disagreement, due mostly to higher eccentricity
and lower numerical resolution (see Fig.~\ref{fig:PhaseComparison}, 
where the larger dephasing due to is apparent). We should note
however that the mismatch is still extremely small, and easily meets
the standard detection criteria, which we will discuss below.

As we pointed out in Section~\ref{sec:waves}, the Fourier transform
of the strain used for match calculations was produced from the
Fourier transform of $\Psi_4$, i.e., we calculated $\Psi_4(t) \rightarrow
\tilde{\Psi}_4(f) \rightarrow \tilde{\hc}(f)$. The matches we
calculate are very close to unity, and we wish to know how much the
results vary if we first calculate the strain $\hc(t)$ in the time
domain, or if we vary slightly the length of the waveforms in either
time or frequency. We tested the robustness of some of our match
calculations to these changes, and found that the results could vary
by as much as $2 \times 10^{-4}$, but the values shown in
Fig.~\ref{fig:matches} were almost always lower than those calculated
by other methods. As such, we consider the curves in
Fig.~\ref{fig:matches} as lower bounds on the mismatch, and note that
we expect that in the worst case they would be no more than $2 \times
10^{-4}$ higher.

The best mismatch required for detection is usually given as 0.035
\cite{Abbott:2003pj,Abbott:2005pe,Abbott:2005qm,Abbott:2007xi},
for which no more than 10\% of signals will be lost~\cite{Owen_B:96}. 
This 
is the best mismatch required between a member of a waveform 
template bank and the true physical waveform that is detected. In 
LIGO detector searches, templates are constructed such that the
worst mismatch between successive members of the template bank 
is 0.03 \cite{Abbott:2003pj,Abbott:2005pe,Abbott:2007xi}. 
This places the more stringent requirement on the accuracy of
the theoretical waveform of a mismatch better than 0.005; see the
discussion in \cite{Lindblom:2008cm}.
This threshold is well above the largest mismatches calculated here. 
The conclusion, then, is that current numerical waveforms 
are sufficiently accurate for detection purposes with all current and
planned ground-based detectors.

\subsection{Parameter estimation} 

We now evaluate the differences between the waveforms with respect to
measurement of the source's parameters. 
The theory of parameter estimation accuracy is developed and discussed
in~\cite{Finn92,Finn_L:1993,Cutler94,Cutler:2007mi}. We defer a detailed
analysis of these waveforms with respect to parameter estimation to future
work, and here make a first analysis based on the criterion proposed 
in~\cite{Lindblom:2008cm}.

Imagine that a GW signal is detected, and the
waveforms studied here are used to estimate the parameters of the
source. Each of our numerical waveforms is slightly different, and if
the detected signal were strong enough (i.e., the SNR were large enough)
the estimated parameters of the source would be different depending on
which numerical waveform we use. However, if the SNR is below a
certain value, any two of our waveforms will be indistinguishable. To
put this discussion in context, a reliable detection requires an SNR
above a threshold which is usually chosen between about 5 and 8,
depending on details of the detector and search performed 
(compare e.g.~\cite{Beauville:2007kp,Abbott:2007xi,Abbott:2008be}),
Therefore, if the SNR has to be below this threshold value 
for two waveforms to be
indistinguishable, then it is meaningless to claim that they agree: we
will never be able to perform an experiment to check. 

\begin{figure*}[tp]
\centering
\includegraphics[width=80mm]{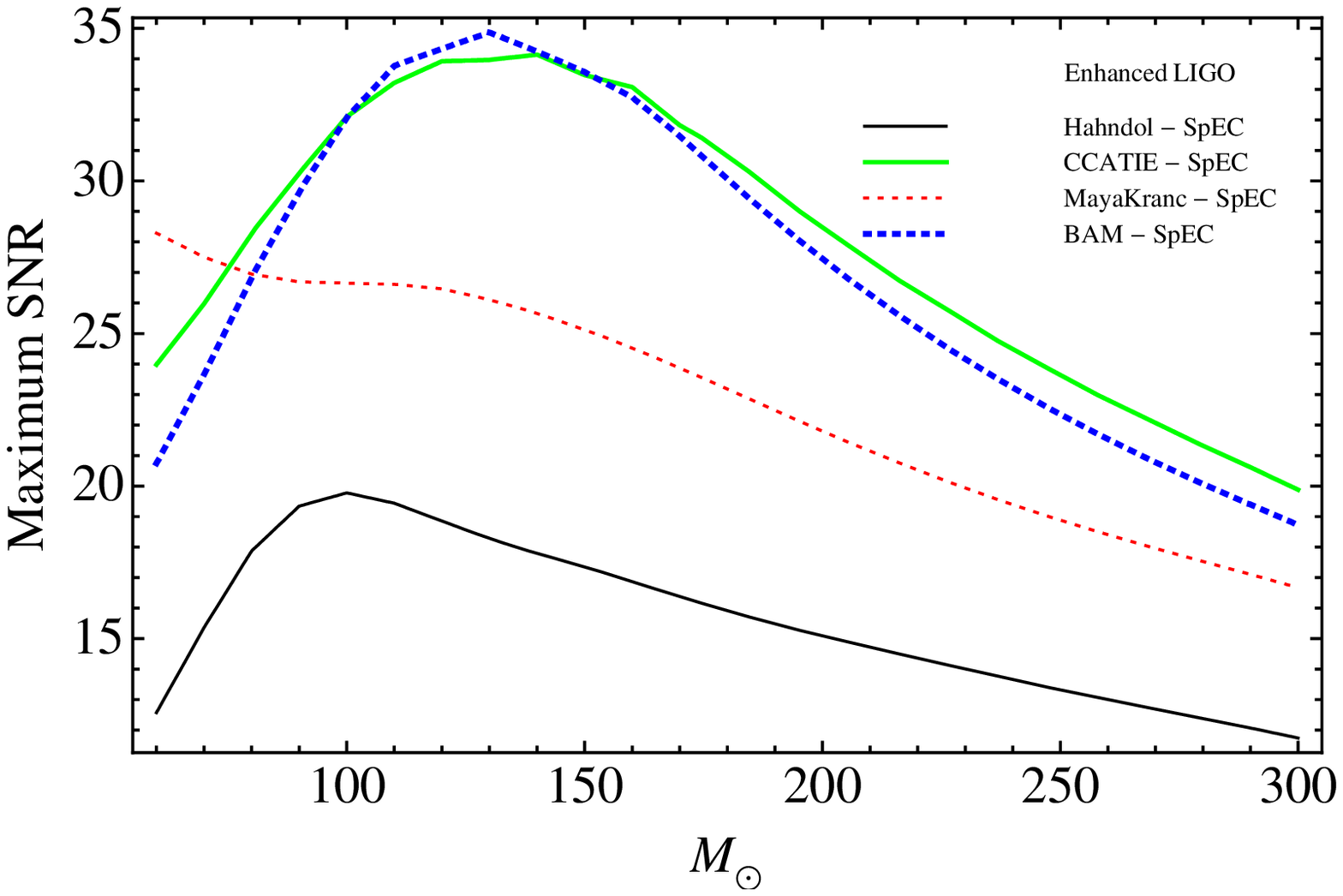}$\quad$
\includegraphics[width=80mm]{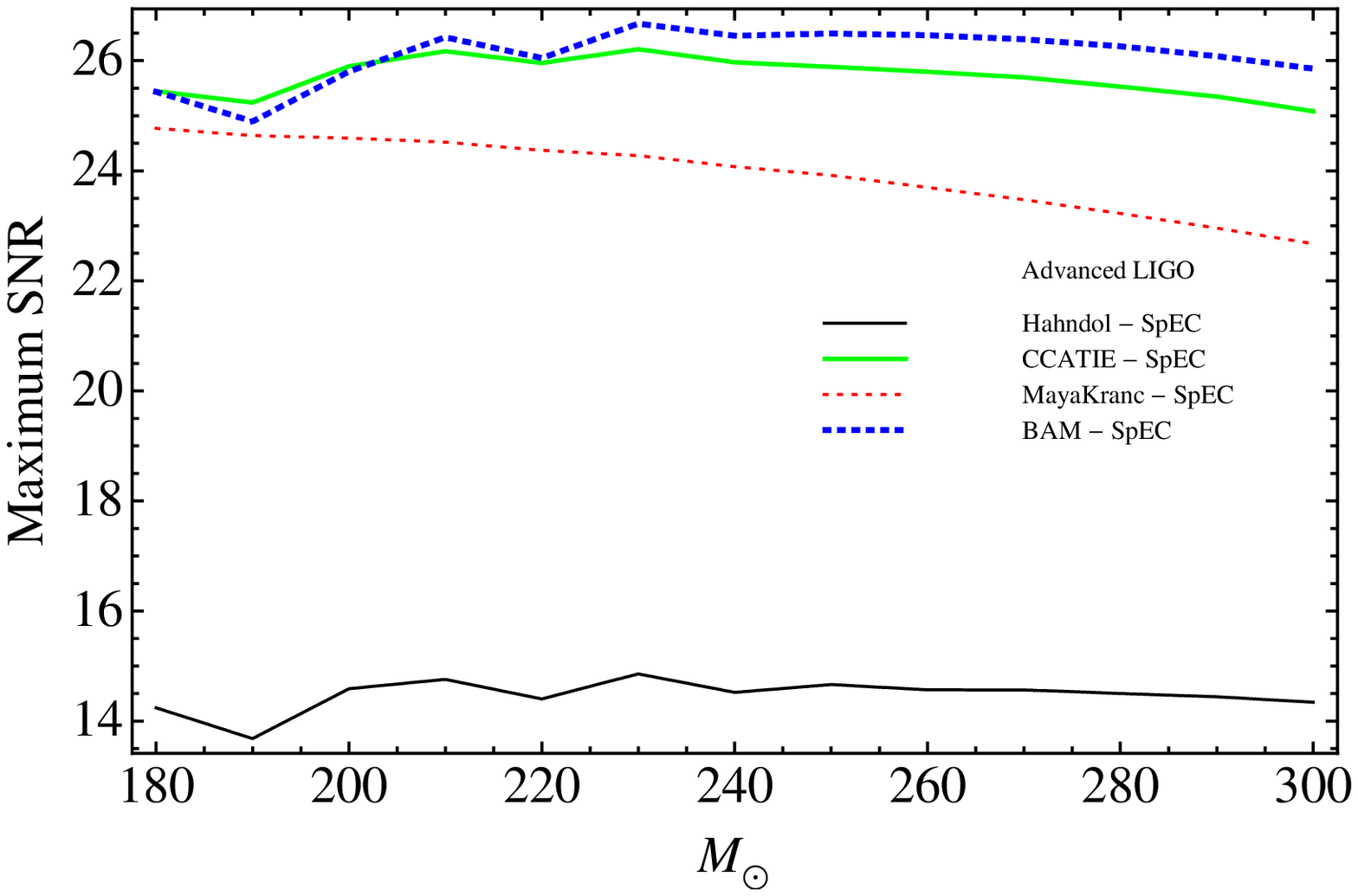}
\includegraphics[width=80mm]{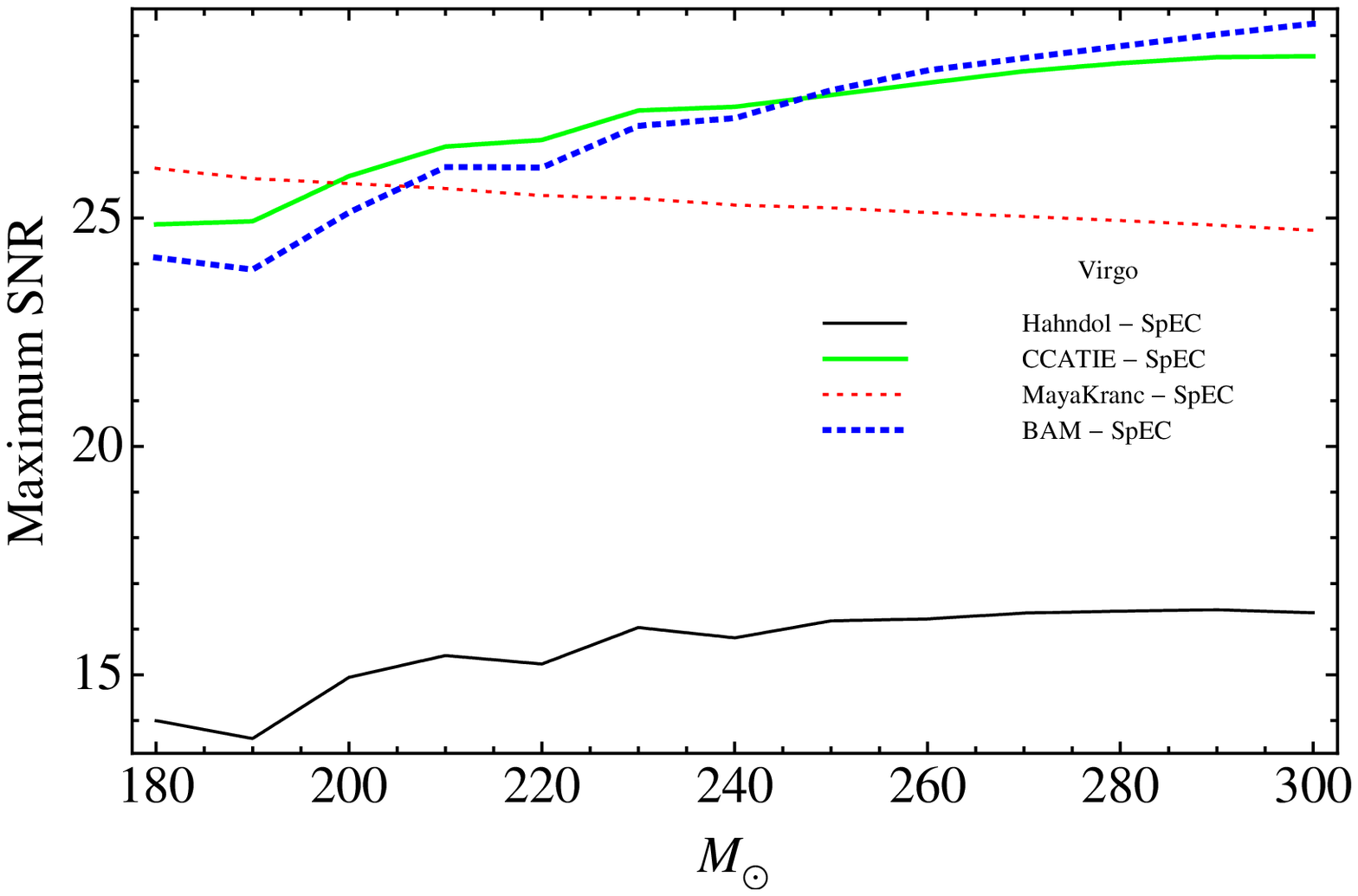}$\quad$
\includegraphics[width=80mm]{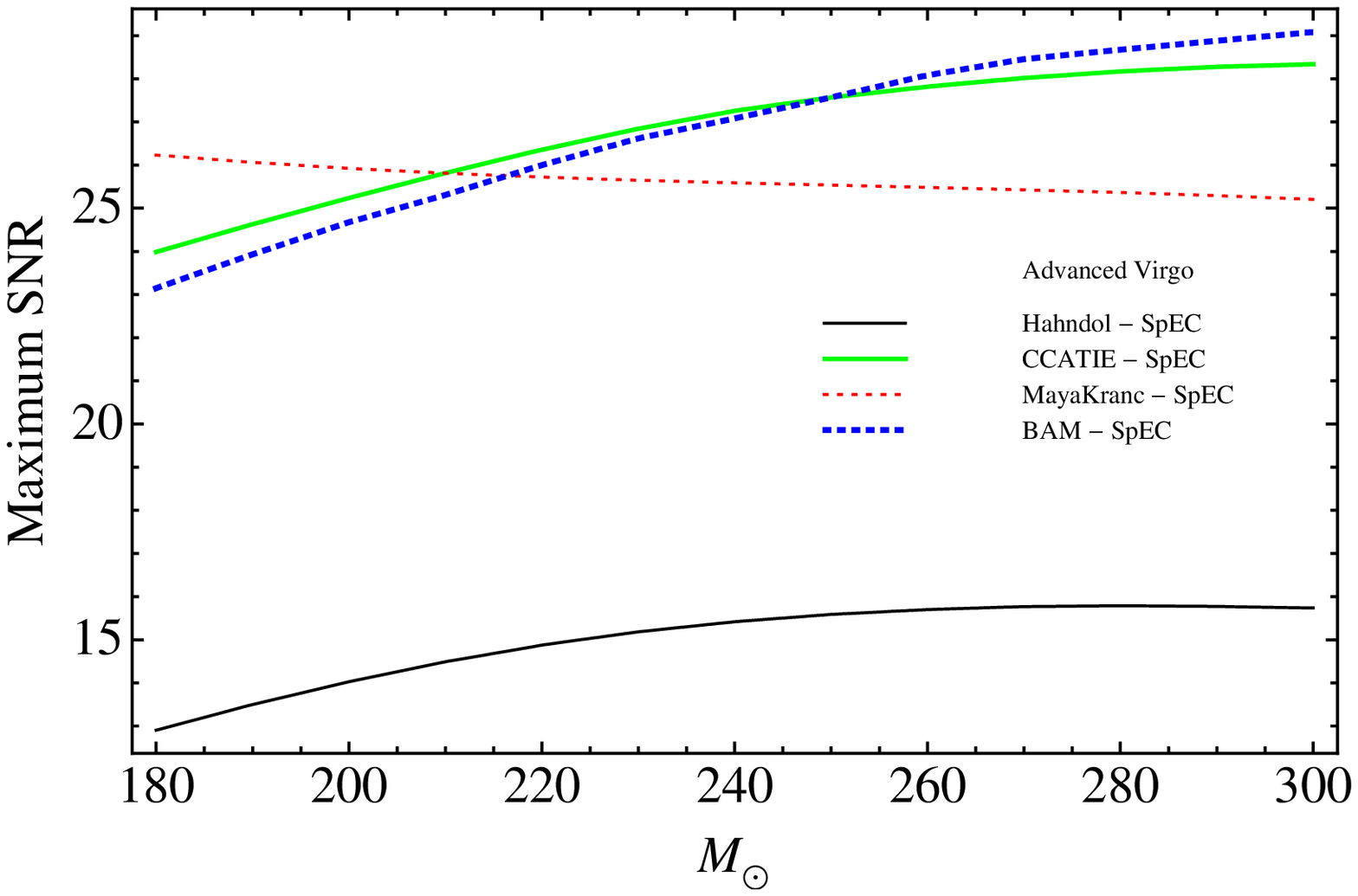}
\caption{The signal-to-noise ratio (SNR) below which the {\tt SpEC} and
  each other waveform will be indistinguishable in any measurement of
  parameters. Results are shown for the Enhanced LIGO, Advanced LIGO, 
  Virgo and Advanced Virgo detectors. See text for further explanation.}
\label{fig:measurement}
\end{figure*}

We can estimate the highest SNR for which two waveforms are
indistinguishable for a single GW detector as follows. Choose the binary mass, the
detector, and distance of the source to the detector. Determine the
time and phase shift such that the mismatch 
between the two waveforms is a minimum. Calculate the difference
between those two aligned waveforms in the time domain, $\delta \hc(t) =
\hc_1(t) - \hc_2(t)$, and transform to the frequency
domain to produce $\delta \tilde{\hc}(f)$.  It was
shown in \cite{Lindblom:2008cm} that when the inner product of $\delta
\tilde{\hc}(f)$ satisfies the criteria \begin{equation}
\langle \delta \tilde{\hc}| \delta \tilde{\hc}\rangle < 1, \label{eqn:measure}
\end{equation} the two waveforms are indistinguishable. The left-hand
side of Eqn~(\ref{eqn:measure}) and the SNR are both inversely
proportional to the distance of the source to the detector. Therefore,
having calculated the SNR and the value of $\langle \delta
\tilde{\hc}|\delta \tilde{\hc}\rangle$ for one source distance, we may
immediately estimate the maximum SNR such that the inequality in
Eqn.~(\ref{eqn:measure}) is satisfied. The results of this calculation
are shown in Figure~\ref{fig:measurement}. 

This analysis applies to any estimation of the {\it intrinsic} parameters 
of the binary, like the total mass and mass ratio. But we should 
emphasize that, since we minimized $\langle \delta
\tilde{\hc}|\delta \tilde{\hc}\rangle$ with respect to a phase and time
shift, Fig.~\ref{fig:measurement} does not apply to an estimation of the 
signal's time of arrival or phase, or parameters that rely on them (like
the sky location). It should also be emphasized that our analysis is
restricted to parameter estimation based on the output of only one 
detector. 

Note also that the relative
performance of each waveform is not necessarily the same for both the
detection and measurement analyses. This is because the match
calculation finds the best agreement in phase only, while the
measurement calculation locates the best match in phase {\it and}
amplitude.

We can estimate from Figure~\ref{fig:measurement}
that the {\tt SpEC}, {\tt BAM}, {\tt CCATIE} and {\tt MayaKranc} waveforms
 will be indistinguishable according to Eqn.~\ref{eqn:measure} if the SNR 
 is below about 25; if the {\tt Hahndol} waveform is to also be
 indistinguishable, the SNR must be below 14. 

SNRs of above 25 are expected to be uncommon for the Enhanced LIGO and 
Virgo detectors; for example, in the NINJA study numerical waveforms were 
injected into simulated detector noises at SNRs no higher than
30~\cite{Aylott08}. 
For the Advanced LIGO and Virgo detectors, however, which have roughly
ten times the range, SNRs in excess of 25 are far more likely.

\section{Conclusion}\label{sec:discuss}

We have compared numerical-relativity waveforms for the last six
orbits, merger and ringdown of an equal-mass nonspinning binary with
minimal eccentricity, as produced by five different computer
codes. We focussed on the $(\ell=2,m=2)$ mode. The purpose
was to perform a stringent consistency check 
of the results from these codes. We verified that accuracy in the
waveform phase and amplitude for each code was consistent with the
uncertainty estimates originally published with each waveform. 

We also
calculated the best mismatch between the most accurate waveform
(calculated with the {\tt SpEC} code) and each of the others, for the
Enhanced LIGO, Advanced LIGO, Virgo and Advanced Virgo detectors, 
and found that it
was below $10^{-3}$ in all cases, and below $2 \times 10^{-4}$ in the 
best cases. Recall that the criteria such that no more than 10\% of signals
is lost is 0.005 (assuming a standard template-bank spacing). 

Finally, we calculated the maximum SNR below which the signals would 
be indistinguishable if observed in the Enhanced LIGO, Advanced LIGO, 
Virgo or Advanced Virgo detectors. For the best cases this is about
25, and is never lower than  
14. This suggests that these numerical waveforms are more than adequate
as ingredients in template banks for GW searches and parameter estimation 
(of intrinsic parameters, at least) with the Enhanced LIGO and Virgo 
detectors, for which an SNR above 25 is unlikely. The {\tt Hahndol}
waveform is distinguishable from the others at an SNR of only 14,
in which case a more detailed study would be
necessary to compare its parameter estimation fidelity with the other
waveforms. Nonetheless, we estimate that {\it less} accurate waveforms
would not be desirable for GW data analysis purposes. We expect that these
results extend to other numerical waveforms, {\it if} they exhibit similar
or better levels of numerical uncertainty. 

An important caveat to the above statements is that these waveforms 
could only be used ``as is'' in detector searches for high-mass binaries. 
Detection of binaries with lower masses would require waveforms that 
are longer (extend to lower frequencies), for example by combining
analytic approximations (usually post-Newtonian and effective-one-body 
[EOB] waveforms) with numerical results. Methods have been proposed
to produce both hybrid waveforms
\cite{Pan:2007nw,Ajith:2007qp,Ajith:2007kx,Ajith:2007xh,Boyle:2009dg}  
and analytic waveforms based on
either a phenomenological ansatz \cite{Ajith:2007qp,Ajith:2007kx,Ajith:2007xh} 
or the adjustment of free parameters in various EOB prescriptions 
\cite{Buonanno:2007pf,Damour:2007yf,Damour:2007vq,Damour:2008te,Baker:2008mj,Mroue:2008fu}.  
It is the accuracy of those ``complete'' waveforms 
that will be important in lower-mass searches, and one may also find that
sufficiently accurate complete waveforms will require either much longer 
numerical waveforms as input, or more physically accurate
approximation techniques. 
Such questions are beyond the scope of this paper, but are an important topic
for future work. 

Furthermore, for general waveforms higher spherical harmonic
modes will play a much more important role, in particular for
parameter estimation. 
This has first been pointed out for ground-based detectors
\cite{Sintes:1999cg} using 
post-Newtonian inspiral waveforms,
with much recent work on ground-based detectors
\cite{VanDenBroeck:2006ar} as well as on the planned space-based  
LISA mission \cite{LISA1,Danzmann:2003tv}, see
e.g.~\cite{Arun:2007hu,Trias:2007fp,Arun:2008zn}.  
Recently significant improvements in parameter estimation for LISA
from higher mode 
contributions have also been hinted at for numerical waveforms
\cite{Babak:2008bu,Thorpe:2008wh}. 
Large values of the SNR will be typical for future generations of
ground based detectors, 
and even more so for LISA detections. This will make it
possible to determine source parameters far more accurately than with current
ground-based detectors, which will in turn place more stringent accuracy
requirements on numerical waveforms. However, we hope that by the time
LISA flies (2018+), and by the time second and third generation ground based 
interferometers are in operation, 
the typical accuracy of numerical waveforms will have 
far surpassed that of those considered in this study. Our more immediate
concern is whether current numerical codes are producing waveforms 
of sufficient numerical and physical accuracy for use in current 
data-analysis applications, and our results suggest that they are, at
least as far as one is  
concerned with the quadrupole mode, which is typically the basis of
current matched-filter searches.

\acknowledgments

The authors thank Badri Krishnan for the Enhanced LIGO noise curve, as 
provided by Rana Adhikari on behalf of the LIGO Scientific Collaboration, 
and Giovanni Losurdo for providing the
Advanced Virgo noise curve on behalf of the Virgo collaboration; 
and Ben Owen and Alberto Vecchio for helpful comments on the
manuscript; and Doreen M\"uller for alerting us to a mislabelling of Fig.~\ref{fig:Phase}. 

M. Hannam was supported by SFI grant 07/RFP/PHYF148, and thanks the
Albert Einstein Institute in Potsdam for hospitality while some of this work was
carried out.
S. Husa has been supported in part as a VESF fellow of the
European Gravitational Observatory (EGO), by DAAD grant D/07/13385  
and grant FPA-2007-60220 from the Spanish 
Ministerio de Educaci\'on y Ciencia.

B. Kelly was supported by the NASA Postdoctoral Program at the Oak
Ridge Associated Universities.
F. Herrmann, I. Hinder, P. Laguna and D. Shoemaker acknowledge the 
support of the Center for Gravitational Wave Physics at Penn State 
funded by the National Science Foundation under Cooperative 
Agreement PHY-0114375. P. Laguna and D. Shoemaker were also supported 
by NSF grants PHY-0653443, PHY-065303, PHY-0555436.
F. Herrmann was also supported by NSF grant PHY-0801213.

J. Baker, M. Boyle, M. Hannam, F. Herrmann, S. Husa, L. Kidder, 
H. Pfeiffer and M. Scheel 
thank the Kavli Institute for Theoretical Physics (KITP) 
Santa Barbara for hospitality during
the workshop ``Interplay between Numerical Relativity and Data Analysis'',
where this work was initiated; the Kavli Institute is supported by NSF grant PHY05-51164.

{\tt BAM} simulations were carried out at LRZ Munich. 
{\tt CCATIE} simulations were supported by Teragrid grant TG-MCA02N014.
{\tt Hahndol} simulations
were carried out using Project Columbia at NASA Ames Research Center. 
Some of the {\tt SpEC} simulations discussed here were produced with 
LIGO Laboratory computing facilities. LIGO
was constructed by the California Institute of Technology and Massachusetts 
Institute of Technology with funding from the National Science Foundation 
and operates under cooperative agreement PHY-0107417.

This work was supported in part by the DFG grant SFB/Transregio~7
``Gravitational Wave Astronomy''; by grants from the Sherman
Fairchild Foundation to Caltech and Cornell, and from the Brinson
Foundation to Caltech; by NSF grants PHY-0601459, PHY-0652995,
DMS-0553302 and NASA grant NNG05GG52G at Caltech; by NSF grants
PHY-0652952, DMS-0553677, PHY-0652929, and NASA grant NNG05GG51G at
Cornell; and by NASA Grant No. O5-BEFS-05-0044 at Goddard.

\bibliography{refs}

\end{document}